%% file: paper.tex
\documentclass[letterpaper,twocolumn,10pt]{article}
\usepackage{usenix-arxiv}

\usepackage{tikz}
\usepackage{amsmath}
\usepackage{graphicx}
\usepackage{enumitem}
\setlist[itemize]{leftmargin=*}
\usepackage{xfrac}

\usepackage{filecontents}
\usepackage{xcolor}
\usepackage{etoolbox}
\usepackage{xspace}
\usepackage{caption}
\usepackage{subcaption}
\newcommand{\mrc}{MRC\xspace}

\newbool{showcomments}
\booltrue{showcomments}

\newcommand{\mycomment}[3]{
    \ifbool{showcomments}{
        {\itshape\small\color{#2}[#1:#3]}
    }{}
}
\newcommand{\htor}[1]{\mycomment{htor}{green}{#1}}

\newcommand{\jijos}[1]{\mycomment{jijos}{orange}{#1}}

\begin{document}

\date{}

\title{\Large \bf Resilient AI Supercomputer Networking using \mrc and SRv6}

\author{
{\bf OpenAI}\\
Joao Araujo\\
Alex Chow\\
Mark Handley\footnotemark[1]\\
Ryder Lewis\\
Christoph Paasch\\
Jitendra Padhye\\
Michael Papamichael\\
Greg Steinbrecher\\
Amin Tootoonchian\\
Lihua Yuan
\and
{\bf Microsoft}\\
S. Anantharamu\\
Abhishek Dosi\\
Mohit Garg\\
Mahdieh Ghazi\\
Torsten Hoefler\\
Deepal Jayasinghe\\
Jithin Jose\footnotemark[1]\\
Abdul Kabbani\\
Guohan Lu\\
Yang Wang
\and
{\bf AMD}\\
K. Doddapaneni\\
Murali Garimella\\
Vipin Jain\\
Yanfang Le\\
H. Nagulapalli\\
S. Narayanan\\
Rong Pan\\
Rathina Sabesan\\
Raghava Sivaramu\\
Rip Sohan\footnotemark[1]
\and
{\bf Broadcom}\\
Eric Davis\\
Dragos Dumitrescu\\
Mohan Kalkunte\\
Bhaswar Mitra\\
Guglielmo Morandin\\
Adrian Popa\\
Costin Raiciu\\
Eric Spada\footnotemark[1]\\
John Spillane\\
Niranjan Vaidya\\
\and
{\bf NVidia}\\
Aviv Barnea\\
Idan Burstein\\
Elazar Cohen\\
Yamin Friedman\\
Noam Katz\\
Masoud Moshref\\
Yuval Shpigelman\\
Shahaf Shuler\\
Shy Shyman\\
Sayantan Sur\footnotemark[1]
}

\maketitle
\footnotetext[1]{Corresponding authors: mjh@openai.com, jijos@microsoft.com, rip.sohan@amd.com, eric.spada@broadcom.com, ssur@nvidia.com}

\begin{abstract}
\input{abstract-arxiv}
\end{abstract}

\input{intro-arxiv}
\input{co-design}
\input{operations}

\input{limitations}
\input{experiments}

\input{relatedwork}
\input{conclusions}
\input{acknowledgments}

\bibliographystyle{plain}
\bibliography{biblio}
\section*{Appendix}
\input{experiments_brcm}

\end{document}

%% file: abstract-arxiv.tex
Tail latency dominates the performance of synchronous pretraining jobs
when running at very large scales.  We describe a three-pronged
approach: (1) a new RDMA-based transport protocol, MRC, sprays across
many paths and actively load-balances between them, eliminating the
issue of flow collisions (2) the use of multi-plane Clos topologies to
get the benefits of high switch radix and redundancy, allowing
training clusters well over 100K GPUs to be built as two-tier
topologies while increasing physical redundancy, and (3) the use of
static source-routing using SRv6 to allow MRC the freedom to
bypass failures by itself.  We describe our experiences running
MRC and static SRv6 routing in production in OpenAI
and Microsoft's largest training clusters, where it has been used to train the latest
frontier models. We demonstrate how MRC allows AI training jobs to
ride out many network failures that previously would have interrupted
training.

%% file: intro-arxiv.tex
\section{Introduction}

As networks for AI training scale into the hundreds of thousands of
GPUs it becomes increasingly difficult to achieve acceptable uptime and
performance for the largest training jobs.  This is especially the case for 
synchronous pretraining, where each step of computation is performed in lock-step by large numbers of GPUs,
interspersed with communication between nodes to perform a combination
of pipeline parallelism, data parallelism, tensor parallelism and
expert parallelism~\cite{10.1145/3320060,rajbhandari2022,yan2026scalable}.  The goal is to overlap communication and
computation to the maximum extent possible, so that the critical path
is balanced and computation-dominated.  Achieving this at scale
is challenging, as the duration of each communications round is 
determined by the slowest transfer.

As computations scale, communication becomes increasingly
outlier-dominated~\cite{10.1145/2408776.2408794} a phenomenon long
known in the HPC community as system or network
``noise''~\cite{10.1109/SC.2010.12,10.1145/1048935.1050204}. To make
matters worse, network failures become more prevalent with scale; when they
cause job failues they become expensive in lost GPU time.  How then
can we maintain acceptable training performance with jobs that require
large numbers of GPUs (up to 100,000 or more) to perform a synchronous
computation?
Broadly,
any solution needs to do three things:
\begin{itemize}
\item Load balance the network evenly, so as to prevent congestion due to flow collisions;
\item Handle incast-based congestion without creating outliers;
\item Handle link and fabric failures gracefully, without bringing down the training job.
\end{itemize}

To add to the problem, very small teams of people need to be able to
manage the networks of multiple supercomputers each with many
thousands of switches running multiple simultaneous training jobs with
their own unique traffic patterns.  We cannot rely on human
intervention to diagnose and fix individual network failures in a
timely manner, so the protocol stack needs to be failure tolerant by
design and the network itself should have a very simple control plane,
which requires almost no active management.  Failed links or
misbehaving switches need to be bypassed automatically.  Failed nodes
however, can already easily be removed from the running jobs, without
requiring global coordination.

To address these issues in very large training clusters we designed,
built and deployed Multipath RC (\mrc), which extends the
RoCE\cite{IBTA:roce} Reliable Connection (RC) semantic layer and draws
upon lessons from Ultra Ethernet Transport (UET)\cite{uet,ue-overview}.  Like UET,
\mrc employs packet spraying, adaptive load balancing based on ECN,
out-of-order memory placement of received data, selective
retransmission, and uses packet trimming to mitigate incast.  Unlike
UET, \mrc is a minimal extension to RoCE; \mrc leverages and extends
the existing Verbs API but our AI workloads only require a subset of
the functionality so, at the transport level, only the RDMA write and
write-with-immediate operations are supported.

Knowing that we would deploy
\mrc allowed us to co-design the topology of training clusters with
very high resilience in mind.  Further, \mrc's adaptive load balancing
is extremely good at routing around failures by itself.  We took the unusual
position of disabling dynamic routing in the switches because we
didn't want two adaptive routing mechanisms interacting with each
other and dynamic routing wasn't adding anything.  Instead data packets are source-routed along static paths
using IPv6 segment routing (SRv6).  Using static routing seems at
first glance to be the opposite of what we wish to achieve, but we
will describe how this combination leads to a highly resilient high
performance training cluster at the largest scale with a low
operational burden.

In this paper we describe our experiences designing, deploying and
operating \mrc at OpenAI and Microsoft.  We implemented \mrc in 400
and 800Gb/s RDMA NICs: NVIDIA ConnectX-8, AMD Pollara and Vulcano, and
Broadcom Thor Ultra.  We also implemented support for SRv6 in NVIDIA
Spectrum-4 and 5 switches running both Cumulus and SONiC, and
collaborated with Arista to implement it in EOS on Broadcom Tomahawk 5
switches.  \mrc is in large-scale production use in multiple very
large AI training clusters, where it has been used to train 
frontier large language models (LLMs) for ChatGPT and Codex. 
We have released the \mrc specification\cite{mrc-spec}
under through OCP under an open license for anyone to use.

%% file: co-design.tex
\begin{figure}[t]
  a)\\
  \vspace{-0.25in}\\
  \begin{subfigure}[b]{\linewidth}
    \centering
    \includegraphics[width=0.92\linewidth]{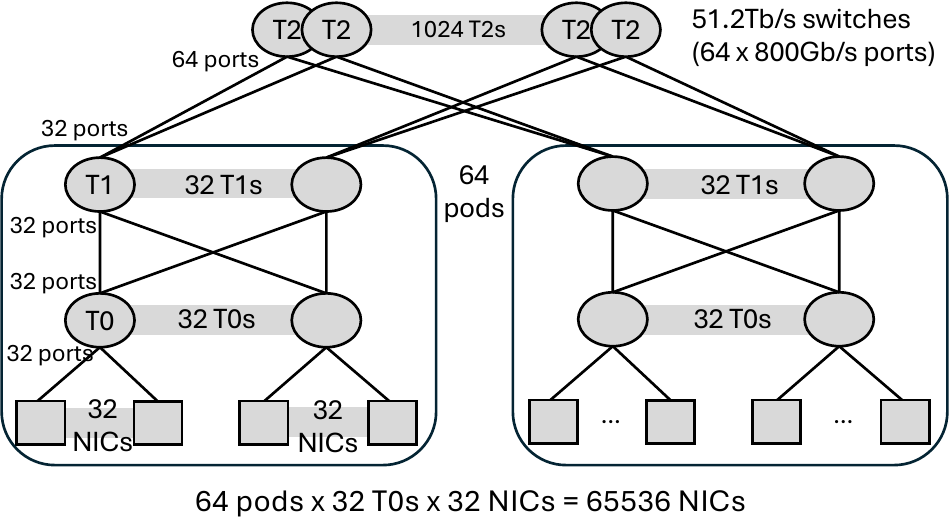}
  \end{subfigure}
  \vspace{0.01in}
  \\b)\\
  \vspace{-0.3in}\\
  \begin{subfigure}[b]{\linewidth}
    \centering
    \includegraphics[width=0.65\linewidth]{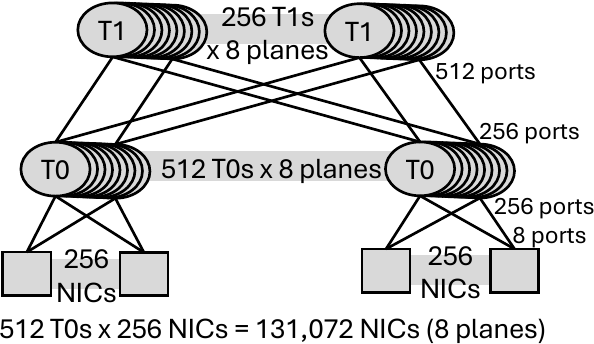}
  \end{subfigure}
  \vspace{-0.2in}
  \caption{(a) 3-Tier 800 Gb/s single-plane topology vs (b) 2-Tier 8x100 Gb/s multi-plane topology}
  \label{fig:topo}
\end{figure}

\section{Multi-plane Topology Co-Design}
\label{sec:topology-codesign}

Consider a hypothetical cluster with 100,000 GPUs, each paired with an
800Gb/s NIC. We wish to achieve full bisection bandwidth to simplify
workload placement. One option is a conventional three-tier Clos
topology using today's fastest Ethernet switches
(Fig.~\ref{fig:topo}a).  Currently datacenter-class switches can
achieve 51.2Tb/s giving 64 ports at 800Gb/s.  Each Tier-0 (T0) switch
connects down to 32 NICs and connects up to 32 Tier-1 (T1) switches,
giving a pod size of 1024 NICs.  Each T2 switch connects to 64
different pods giving a cluster size of 64K NICs.  If we wish to
connect 100K GPUs we either need to use four tiers of switches,
oversubscribe the network, or build multiple independent rails.

Alternatively, we can break out the 800Gb/s NIC by lane and use it as
8 x 100Gb/s ports (Fig.~\ref{fig:topo}b).  We build eight parallel 100Gb/s Clos
planes using the same 51.2Tb/s switches, but now each switch has 512
ports.  Each T0 switch connects down to 256 NIC ports and connects up to 256 T1
switches.  Each T1 switch connects in turn down to 512 T0 switches, giving
a network of 131,072 GPUs.  In this arrangement we can easily
accommodate our 100K GPUs using {\em only two tiers of switches}.  Such a
multi-plane network has many advantages: 
\begin{itemize}
  \item Latency is lower because the longest path traverses only three
    switches rather than five or seven.
  \item Many more nodes are reachable in one hop (256 rather than 32),
    so it is easier to take advantage of locality in the job, reducing
    latency and reducing load on T0 uplinks.
  \item Cost and power consumption are reduced - for full
    bisection bandwidth we require $\sfrac{2}{3}$ of the optics and $\sfrac{3}{5}$ the number of switches compared to a 3-tier network.
  \item The impact of an in-network failure is much less.  For
    example, losing a T0-T1 link reduces capacity from a node by ~3\%
    in an 800Gb/s plane, vs ~0.4\% in a 100Gb/s plane.
  \item It is possible to lose a NIC-T0 link without bringing down the
    training job.  We still lose 12\% of the NIC bandwidth, but can
    easily ride out a link flap with the remaining capacity.
\end{itemize}
It seems clear, therefore, that a multi-plane design has many
advantages over a single-plane design, but there are some challenges
too. First, the workload needs to be able to survive link and NIC port
failures.
Second, to fully use the network we need to be able to load
all network planes equally and load balance across all the many paths
within a plane without suffering a loss of performance due to flow
collisions.  This is hard to do with traditional single-path transport protocols~\cite{hedera}
and is made harder by using lower speed network links which are easier
to overload.  This is where \mrc comes in.

\subsection{\mrc Overview}

\mrc extends the RoCEv2 Reliable Connection (RC) transport protocol to
support multi-path operation borrowing several features of UET\cite{uet,ue-overview}.  It supports the normal RoCE verbs
interface and Queue Pair (QP) abstraction, but only for write and
write-with-immediate operations.  At a protocol level, the main
features \mrc adds are the following:
\begin{itemize}
\item Every data packet contains the RDMA virtual address and remote
  key so the receiving NIC can write each arriving packet
  to memory immediately, no matter the arrival order.
\item Each packet contains an entropy value (EV) that dictates its path
  through the network.  The 32-bit EV is striped across the UDP source
  port and IPv6 flow label in an \mrc packet.  In a conventional
  network, changing the EV causes switches to hash each packet to a
  different path from the ECMP set.  At QP startup, the sender generates
  an EV set for that QP---typically 128 to 256 entries. The sender then
  rotates through this set, using a different EV for each packet, so
  that all packets of a QP are sprayed across many paths on all planes in a multi-plane network without the application needing to know.  This serves
  to load balance the network.
\item Spraying is hard to combine with the priority flow control (PFC)
  mechanism used in lossless Ethernet because a single flow reaches
  the last-hop switch over hundreds of paths. Further, PFC tends to
  create head-of-line blocking between different collectives, hurting
  tail latency.  Thus \mrc disables PFC and uses Ethernet in
  best-effort (lossy) mode.
\item The combination of best-effort Ethernet and out-of-order
  delivery places a greater burden on recovering losses quickly.  \mrc
  implements fast selective retransmission, using Selective ACK (SACK)
  packets to indicate precisely which packets have arrived at the
  receiver.
\item To further increase retransmission speed, especially under
  incast, MRC can use packet trimming~\cite{ndp,uet}. With packet trimming, a
  packet that would have been dropped due to congestion has its
  payload trimmed off and is priority-forwarded to the
  destination. The receiving NIC then generates a NACK to trigger fast
  retransmission. This also lets MRC distinguish congestion loss from
  other packet loss, which in AI clusters is mostly due to link flaps
  and failures.
\end{itemize}

A protocol like \mrc, designed around packet spraying, is a
very good fit for a multi-plane network.  Each EV corresponds to a
specific path on a specific network plane.  When \mrc generates its
EV set, it chooses an equal number of EVs per plane.  This
immediately equalizes the traffic between planes.

For each EV, \mrc keeps a few bits of state about path health.  In each switch,
we enable Explicit Congestion Notification (ECN) in the normal
randomized manner, but disable ECN on the last hop to the receiver. In
a network with full bisection bandwidth, the traffic aggregate should
not experience congestion, except from incast on the last hop, so ECN
now acts as a load-balancing signal. The receiver echoes the ECN
signal back to the sender, indicating that this specific path is more
congested than others, and the sender temporarily avoids it. Different \mrc
senders do not coordinate when choosing their EV sets,
so even though each sender load balances well, the aggregate may be
slightly uneven. ECN-based load balancing smooths out this unevenness,
keeping internal queues from growing enough to cause congestive loss.

When a packet is not trimmed but actually lost, \mrc assumes the path
has failed and immediately stops using the corresponding EV.
Of course, not all loss is due to failed paths - packets can suffer bit
errors or other issues - so permanently retiring an EV after one lost
packet may leave us short of working EVs. To avoid this, \mrc sends background
path probes to determine whether paths it assumed were bad are actually bad, and also detect if failed links have recovered. If
enough probes succeed, the EV is resurrected.  

At this point we have a transport protocol that can detect path
failures and bypass them in a few tens of microseconds.

\subsection{Static Segment Routing}

In a conventional datacenter network we would use a dynamic routing
protocol such as BGP to determine reachability and route around failed
links. Generally this works, though convergence can take a very large
number of RTTs.  Unfortunately, high-radix two-tier topologies make
this worse. Every destination corresponds to a large ECMP set - up to
256 entries in a 512-port T0 switch, one per uplink. This is not a
problem without failures, but most T1 switches may have at least one
failed downlink. As a result a specific destination is unreachable via
some T1 switches, so a T0 switch cannot use the default ECMP set
balanced across {\em all} uplinks to reach it. The number of large
ECMP sets required in each T0 switch grows close to the total number
of T0 switches. Maintaining so many large ECMP sets can be challenging
for dynamic routing and for switch forwarding engines to
support. Diagnosing routing problems at scale is notoriously
difficult.

Given \mrc will rapidly avoid failed paths and that we build
redundant topologies, do we still need dynamic routing?  In fact, combining
end-system load balancing with switch-based dynamic routing makes
network behavior harder to understand than either running alone. MRC
reacts to a failure first, avoiding the broken path; then
dynamic routing re-routes, changing ECMP mappings and disturbing load
balancing.

We took the position that, when running MRC, dynamic routing caused more
problems than it solved, so we simply disabled it, but we still need a way
to map EVs to specific network paths. One option would be to
run MRC with statically configured ECMP routes, relying on MRC to avoid
bad paths. The problem is that there still is not a simple mapping from
EV to path. For example, it would be hard to correlate MRC's EV-based
view of bad paths with the precise physical path, so
we can report failures for repair. This led us to spray using source
routing, as prior work has suggested~\cite{ndp}.

The approach we chose was to deploy IPv6 segment routing (SRv6)~\cite{rfc8986}.  In
the \mrc NIC, at QP startup a set of entropy values (EVs) are chosen,
such that bits in each EV {\em directly embed} the path choice available at each hop in the
network.  Each EV then maps to a specific unique path through the
network to the destination NIC.

SRv6 is a family of solutions;  we use the
micro-segment ID (uSID) format~\cite{rfc9800}, where the destination IPv6 address
consists of a 32-bit locator prefix followed by a sequence of 16-bit
uSIDs each corresponding to a specific switch along the path.  We use
the uN style of uSID, where each switch along the path is explicitly
named. We describe the relation between EVs and SRv6 addressing in detail
in Section~\ref{evs}.

\begin{figure}[t]
  \centering
  \includegraphics[width=0.95\linewidth]{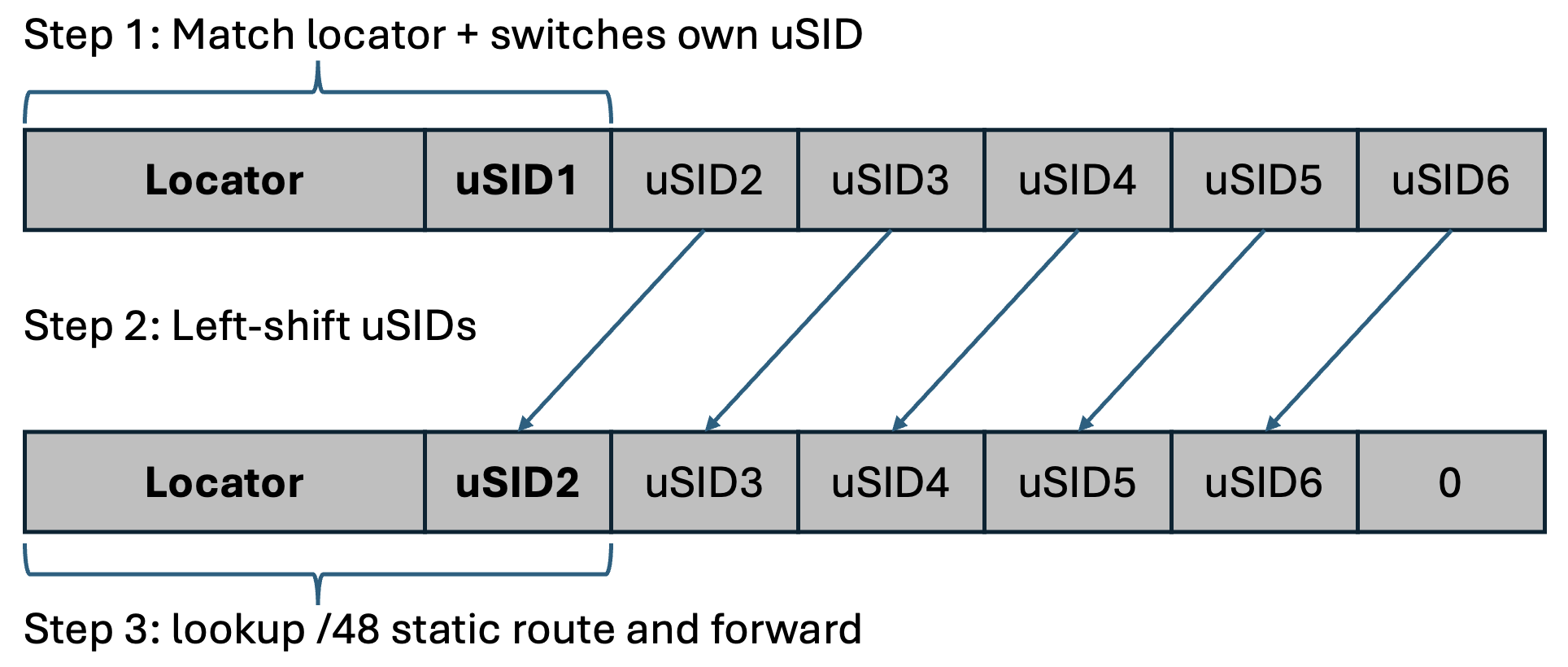}
  \caption{SRv6 forwarding using uN uSIDs}
  \label{fig:srv6}
\end{figure}

The SRv6 forwarding process is shown in Fig.~\ref{fig:srv6}.  When a
packet arrives, the switch compares the first 48 bits of
the destination address with the switch's configured SRv6 locator and uSID; if they match this is a valid SRv6 packet for the switch to
process.  The switch then left-shifts the uSID part of the address by
16 bits, moving the next-hop uSID into the first 48 bits.  This new
address is then looked up in the switch's static forwarding table in the
normal way. This forwarding table was configured when the switch was installed, and
is generally never changed.  The static route dictates which egress port
the packet uses.  In all switches we deploy, this uN style of SRv6
forwarding can be performed at line rate.

\mrc packets are IPv6 in IPv6 encapsulated, with the outer destination
address being the SRv6 path and the inner destination address
containing the destination NIC's own address, so that the receiving NIC
can recognize and decapsulate the packet before feeding it to the MRC
RDMA pipeline.

\subsection{Mapping EVs to SRv6 Addresses}\label{evs}

MRC was designed to work with either hash-based ECMP forwarding or SRv6.  The
EV is embedded in each packet, striped across the UDP source
port and the IPv6 flow label, both of which are hashed by switches
performing ECMP forwarding.  The EV is also echoed in SACK and NACK
packets to indicate the congestion state on a path.

When using SRv6, although the EV is not hashed by switches, it still
needs to be carried in data packets so the receiver can echo it.  The
SRv6 address itself cannot be echoed, as it is erased by the shifting
process during forwarding.

To avoid holding both an SRv6 address and EV state per path, we use an
algorithmic mapping between each EV and its corresponding SRv6
address. Switch uSIDs are allocated according to the network
structure, allowing the EV value to be a compressed representation of
the bits that vary between SRv6 paths to that destination, plus the
NIC port to be used.

\begin{figure}[t]
  \centering
  \includegraphics[width=0.95\linewidth]{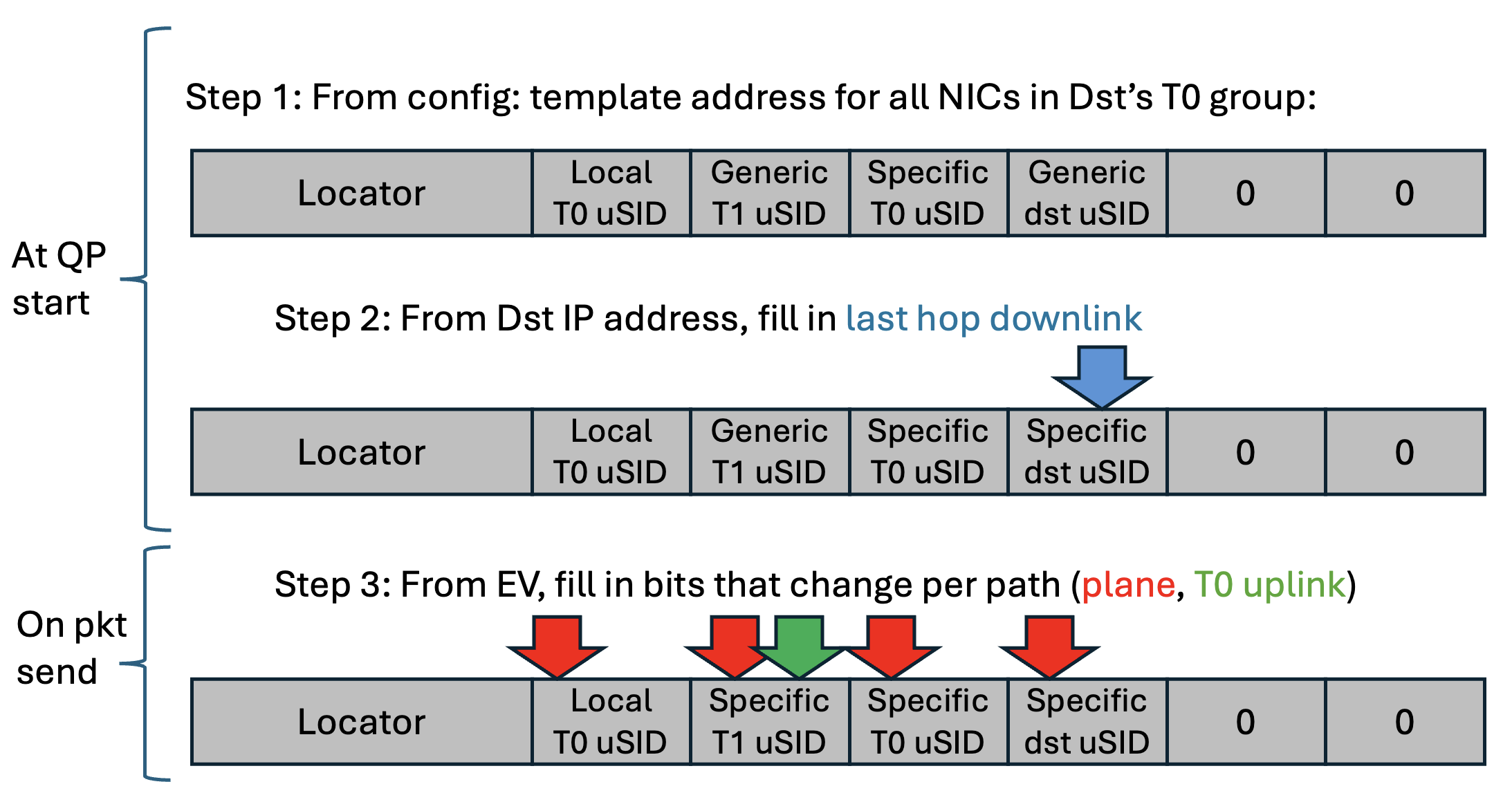}
  \caption{Creating the SRv6 address from an EV and template}
  \label{fig:srv6-addr}
\end{figure}

Fig.~\ref{fig:srv6-addr} illustrates how SRv6 destination addresses are
created for a QP traversing
src$\rightarrow$T0$\rightarrow$T1$\rightarrow$T0$\rightarrow$dst.  On QP
startup, the NIC looks up the destination address prefix in a
node-specific configuration file to obtain a generic SRv6 address
template for nodes in that row. The dst uSID in this template is
specialized for this destination by copying in the last-hop downlink
number. This template is used by all packets sent by the QP.

The NIC also creates a set of EVs for this QP, striping across planes
and paths within each plane.  Each time a packet is sent, a new EV is
selected from the active set.  The template is then further specialized
to create the final destination address by copying the plane number from
the EV into all uSIDs and the T0 uplink number into the T1 uSID.  We
also use an extension of this scheme to forward between
clusters.

\subsection{Choosing Working Paths}

With a conventional protocol, resource management is split between
routing and transport: routing delivers working paths to transport,
and transport manages congestion on those paths.  In a statically
source-routed MRC network, all resource management and failure
handling is the role of MRC.

On QP startup, MRC must select a set of EVs, each mapping to a unique
path on a specific plane.  MRC chooses EVs equally split across the planes, then randomly
selects a subset of paths within each plane. Different senders in the
same T0 group do not coordinate their choices.

At pretraining job startup we ensure all nodes in the job have all
NIC ports operational. MRC supports a denylist, allowing us to avoid paths
that traverse links known to have failed. To enable this, we implemented
Clustermapper consisting of an agent on every node; together these map which links are
currently down or have excessive loss. Static SRv6 routing makes this
very simple: unlike with ECMP hashing, we know exactly which path a
Clustermapper probe packet will take, and know it is the same path an
equivalent MRC packet will take. Unlike switch-based telemetry, the
resulting map gives ground truth about forwarding-plane
health.

\begin{figure}[t]
  \centering
  \includegraphics[width=0.85\linewidth]{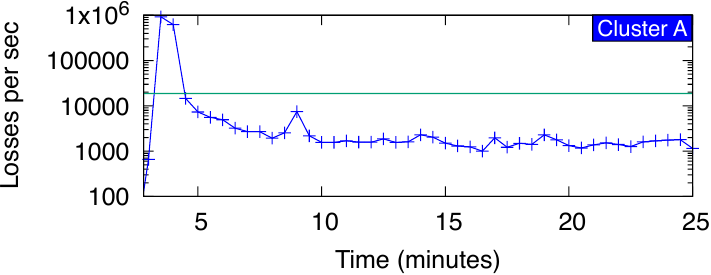}
  \captionsetup{skip=3pt}
  \caption{Startup losses without mapping out bad paths}
  \label{fig:startup}
\end{figure}

In fact, for pretraining it has not proved necessary to use
Clustermapper to set denylists for failed T0--T1 links. QPs are very long
lived and MRC quickly routes around failures and settles on good
paths. On CX8 MRC each QP starts by populating a large EV
set, typically over 100 entries, plus a backup EV set for
failures. The QP starts spraying packets across the corresponding
paths. Some paths are down; packets are lost and retransmitted, and
the EV is swapped with one from the backup set. Even without prior
knowledge, MRC maps out bad paths quickly enough that the slight
startup performance loss is not a problem.

Fig.~\ref{fig:startup} shows the packet loss rate as a 75K GPU
pretraining job starts without pre-populating the denylist. The figure
shows the total losses per second on one of the four NICs per node,
summed across all nodes in the job. The graph uses a logscale to make the
background rate visible. The horizontal line shows one loss per
second per NIC - at 800Gb/s this would be a loss rate of 1 packet in 25 million. The loss
rate across the whole job falls well below this point within a couple
of minutes, but even in the first minute less than 5 packets per QP
are lost. Given that we have to ramp up large training jobs slowly to
avoid destabilizing the power grid, this startup transient has minimal
impact on training time.

\paragraph{\bf Reverse Paths:}
MRC sends RoCE cumulative ACKs, MRC SACKs and MRC NACKs on the reverse
path.  The EV set on the forward path is updated based on SACK and
NACK information, but what EV should reverse path packets use?  MRC
isn't especially sensitive to reverse-path loss, but it does have some
impact on tail latency.

With bidirectional traffic, control packets could use any EV from that
endpoint's active EV set, but many collective QPs are unidirectional at any
instant. We could reverse the SRv6 forward path, but unfortunately this
won't always work.

One solution we have found to work well is to keep a small reverse EV
set for control packets, with at least one EV per plane. Each RTT that
the QP is active inbound but generates no outbound traffic, the
receiver sends an EV probe packet using a randomly chosen EV. If the
probe is acknowledged, the reverse EV for that plane is set to the
probe's EV. If data traffic is sent, the reverse EV is updated from
data SACKs instead of needing EV probes. Thus the reverse EV set
always contains EVs known to be up and working.

%% file: operations.tex
\section{Operations}

A goal of \mrc was to simplify network operations, so that
very large supercomputer networks can be operated by small teams.  Key
to this is that with \mrc, most network failures do not require quick
reactions from the network operators.

\begin{figure}[t]
  \centering
  \includegraphics[width=0.75\linewidth]{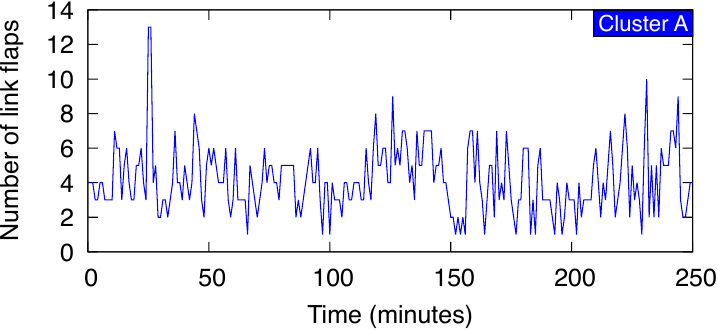}
  \caption{Link flaps between T0 and T1}
  \label{fig:flaps}
\end{figure}

Our experience has been that link failures and flapping links between
T0 and T1 switches can largely be ignored. We still schedule failed
links for repair, but at lower priority. Figure~\ref{fig:flaps} shows
the total number of link flaps per minute between T0 and T1, as
reported by the switches, while running a very large synchronous
pretraining job on Cluster A (Table~\ref{tab:exp-setup}). These links are left
in service because they simply do not cause a problem. \mrc spreads
traffic across enough paths that when a link used by a QP flaps, only a
very small number of packets per QP are lost, often just one, and the
corresponding EV is removed. The missing packet is selectively
retransmitted on a different path, and the impact on the job is
negligible.

\begin{table*}[t]
\centering
\caption{Experiment Platform Configurations}
\vspace{-0.1in}
\label{tab:exp-setup}
\begin{tabular}{l l l l}
\hline
\textbf{Cluster} & \textbf{NIC} & \textbf{Switch} & \textbf{Topology} \\
\hline
	Cluster A & NVIDIA GB200 + CX8 (800 Gbps) & NVIDIA SP4 \& BRCM TH5 & 2-Tier 4 x 200 Gbps Multi-plane \\
	Cluster B & NVIDIA GB200 + CX8 (800 Gbps) & NVIDIA Spectrum 5 & 2-Tier 8 x 100 Gbps Multi-plane \\
	Cluster C & AMD MI355 + Pollara (400 Gbps) & Broadcom Tomahawk 5 & 2-Tier 4 x 100 Gbps Multi-plane \\
	Cluster D & NVIDIA RTX 6000 + Broadcom Thor Ultra & Broadcom Tomahawk 5 & 2-Tier 400 Gbps Single-plane \\     \hline\hline
\end{tabular}
\vspace{-0.2in}
\end{table*}

Prior to \mrc, when a flappy link was banned for repair, it would be
administratively taken down so data traffic could not use it. Only once
it was repaired and tested would it be brought back up. Initially we
took this approach with \mrc too, but it became clear it was
unnecessary. Now we leave links in service while they are being
repaired: \mrc maps them out when they drop, and only brings them back
when enough probes succeed over time. This requires less
coordination and ensures links are available whenever they actually
work, but not used when too unreliable. It is also robust to the
inevitable disruption caused when a technician repairing one link
disturbs neighboring links, causing them to flap.

Switch software is complex and so prone to
bugs\cite{switchfailures,bugs-os-nos,crystalnet} which can hinder fault
detection and sometimes cause the control plane to fail to route around
failures. According to a datacenter study covering 180,000 switches over
three months \cite{switchfailures}, 17\% of switch failures are due to
software bugs. Such bugs can cause control-plane and
dataplane divergence, where the control plane fails to accurately detect
dataplane failures \cite{crystalnet,switchfailures}.
These
studies match our own experience: we have seen a wide range of bugs,
including cases where the control plane and links are up, but the switch
stops forwarding packets. This last case is particularly pernicious.

When using static SRv6, \mrc doesn't care about the state of the
control-plane: if packets don't flow, it removes the path. When we see
a T1 switch that is not happy, we simply reboot it without concern for
routing convergence or needing to coordinate with active training
jobs. \mrc maps out EVs that traverse the failed switch and restores
them afterwards, with negligible effect on job performance.  We do
have to be more careful with NIC-T0 links and T0 switches, as these do
have a measurable effect.  A failed NIC link will not cause a QP
failure; rather the NIC detects the link has dropped and \mrc remaps
EVs to avoid the failed port.  Many packets are lost right after the
link drops.  Remapping all the EVs from the many QPs in use is not
instantaneous in the CX8 \mrc implementation, so this causes a glitch
in job throughput.  \mrc then uses a port state bitmap in SACK packets
to notify remote QP endpoints that the port is down, so they also
remap their EVs to avoid the failed plane.  Once this has happened,
which typically takes a few seconds, the QPs are back to being fully
functional, but using one fewer plane.

We have deployed both four-plane (4 x 200Gb/s) and eight-plane (8 x
100Gb/s) MRC supercomputers. The impact of a NIC port failure is
obviously smaller with eight-planes, but even with four planes the
effect on training job performance can be relatively small -- the
precise amount depends on job layout, but can be significantly less
than the lost capacity. We detect the failure and allow the job to
continue with reduced performance. Most of the time the port recovers
relatively quickly with no lasting impact.  When a port fails down and
stays down, we ban the node and report it for repair.

Not requiring control-plane action for a training job to ride out
failures is only part of the operations story. We still need good
telemetry to root-cause failures, tune job performance, and of course
debug \mrc itself.

Clustermapper is key here. The Clustermapper agents running on all
cluster nodes together probe every link in the network every
millisecond. This gives us fine-grained health data, allowing us to
schedule links for repair or switches for rebooting. A Clustermapper
agent on each node probes each of the 16 or 32 directly connected T0s
(one per port, on each of the four NICs per node) by sending probes
source-routed to the T0 and back to the same agent. This identifies
issues with NIC--T0 links or T0 switches. Each agent also probes a
subset of the T1 switches, again source-routing to the T1 and back, so
that all T0--T1 links are probed at high frequency. The combination of
T0 and T1 self-probes lets us immediately localize precisely which
link or switch has a problem: if T0 probes show no problem but T1
probes do, we know the issue is the T0--T1 link, not the NIC--T0
link. We could run this probing on demand, as soon as \mrc reports a
path problem, but in practice we found it useful for Clustermapper to
detect issues even when no workload is running, and the cost of
continuous probes is low.

Without SRv6, it is harder to get this level of ground truth on
network health. It is much easier to interpret data from probes sent
from an agent through the network back to itself, rather than
mechanisms like pingmesh\cite{pingmesh}, which must send to a remote
node that may not be up. Unlike with ECMP hashing, there is no
ambiguity about which path a probe packet takes, there is no dynamic
routing changing forwarding paths underneath, and we know \mrc data
packets take the same paths.  While directly pinging switches is
possible without SRv6, ICMP probes are handled by the control plane,
which limits probing frequency. With SRv6, the switch treats
a probe like any other data traffic, handling it in the dataplane
and enabling high-frequency probing.

%% file: limitations.tex
\section{Inter-plane Loading }

We took two decisions to simplify MRC behavior that have consequences.
First, when MRC maps an EV out of its active set, it replaces it with
one from the same plane. This ensures load balancing remains exactly
equal between planes when all NICs have all planes available. We made
this choice to avoid false incast at the destination, where mild
congestion on different flows' T0 uplinks can cause them to load
planes unevenly, making some planes more congested than others when
flows converge at a destination.

Keeping the active EV set equal between planes means two cases are not
handled well by default. First, if any single-path traffic is present on
the back-end MRC network, MRC will be constrained by the most congested
plane and lose capacity. This is obviously not a problem if the back-end
network only runs MRC.
A consequence is that if we lost sufficiently many T0--T1 links in a
single plane, that plane would become the bottleneck and we would lose
performance. In practice, we have not seen this in production.

Second, and partly as a consequence of keeping planes evenly loaded, if
a NIC--T0 link does not fail completely but instead has an unacceptably
high packet loss rate, MRC cannot rebalance to avoid the bad plane
because it cannot reliably tell whether the problem is at its end or the
remote end. Clustermapper can, because it probes to the local T0 and
back. Thus we delegate detecting this case to a Clustermapper policy
decision; the bad plane can then be avoided by specifying a matching
denylist entry.

In practice, keeping all planes evenly loaded is a very useful
invariant. All planes normally look the same in network stats once MRC
has avoided bad links, so if one plane looks worse than the others, it
generally points to a network problem.

%% file: experiments.tex
\section{Experiments}

\input{experiment-setup}

\input{experiments-oai-arxiv}

\subsection{Testbed Results}

We present a range of
experiments that demonstrate how MRC behaves in more controlled testing
environments using the MRC implementations on NVIDIA's CX-8 NIC, 
AMD's Pollara NIC and Broadcom's Thor Ultra NIC.

\input{experiments_nvidia}
\input{experiments-amd}

\input{experiments-collateral}

%% file: experiment-setup.tex
\label{sec:exp-setup}

We present results from four different AI Training clusters; the cluster
configuration and topology details are presented in Table \ref{tab:exp-setup}.
All the clusters follow the 2-Tier multi-plane topology as described in 
Figure \ref{fig:topo}b. Each multi-plane NIC connects to a group of T0 switches,
one per plane. T0 switches of corresponding planes are merged at T1 layer.

If a flow goes from
one NIC to another NIC in the same T0 group, we refer to it
as T0-local. Conversely, if the endpoints are in different T0
groups, the flow must traverse the T1 switches; we refer to it
as cross-T1.

%% file: experiments-oai-arxiv.tex
\subsection{Training Results}

MRC has been used to train OpenAI's most recent frontier models at very
large scale. The constant rate of T0--T1 link flaps in Cluster A,
shown in Fig.~\ref{fig:flaps}, has had almost no impact on
performance.  In fact, fixing these flaps is a very low priority
activity, as the operational effort is better spent elsewhere.

We have observed many back-end network failures during training jobs;
very few cause job failure or significant performance degradation.
Losing NIC--T0 links does affect performance, though we observe that
most such events are transient and the job can often recover to full
speed quickly without evicting nodes.

\begin{figure}[t]
  \centering
  \includegraphics[width=0.95\linewidth]{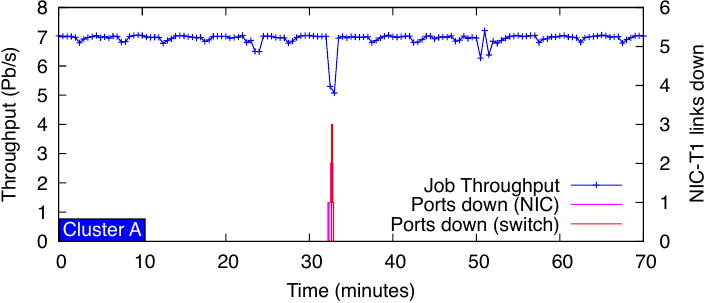}
  \caption{Impact of a flapping NIC-T0 switch transceiver}
  \label{fig:nic-flaps}
\end{figure}

Fig.~\ref{fig:nic-flaps} shows an event that occurred during a 50K
GPU production pretraining job at OpenAI using MRC on CX-8 NICs in
Cluster A.  This supercomputer network is
configured as 4 x 200Gb/s ports per NIC, with MRC spraying each QP
across all four ports.  An optical transceiver on a T0 switch suffered
a glitch, and flapped all its four links in rapid succession. These
four links connected to NICs on four different nodes, of which three
were active in the training job at the time.  The graph shows the
times the NICs think the link was down and when the switches think
the link was down, which are not precisely the same times.

\begin{figure}[t]
  \centering
  \includegraphics[width=0.95\linewidth]{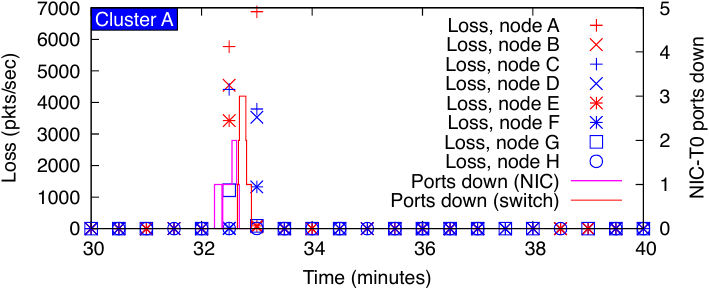}
  \caption{Packet loss rates during the event in Fig.~\ref{fig:nic-flaps}}
  \label{fig:nic-flap-losses}
\end{figure}

As this is a synchronous pretraining job, the slowest node dictates
overall performance. In Fig.~\ref{fig:nic-flaps}, throughput suffered
approximately a 25\% reduction over the minute of flaps, then recovered
to full speed immediately afterwards. The job did not crash, no QP
failed, and the affected nodes did not need to be removed from the job.
Such transceiver glitches are rarer than individual link flaps, but they
do happen occasionally.

Fig.~\ref{fig:nic-flap-losses} zooms in on the loss rates of the
eight most affected nodes. Red nodes are those whose ports
flapped, whereas blue ones were actively sending to the
affected nodes at the time of the flap. MRC recovered the losses quickly
enough that the impact on the job was relatively small.

\begin{figure}[t]
  \centering
  \includegraphics[width=0.95\linewidth]{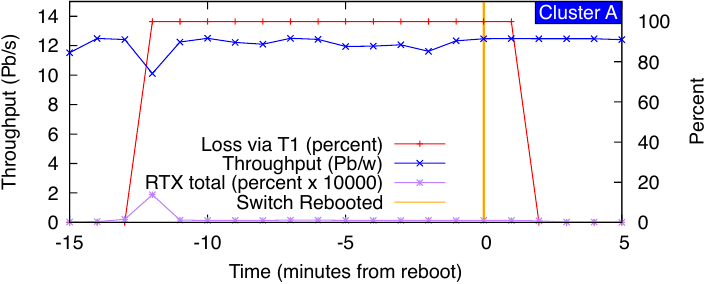}
  \caption{Impact of a T1 switch failure and reboot}
  \label{fig:t1-reboot}
\end{figure}

During a 75K GPU pretraining job, we had four cases where T1 switches
had to be rebooted. Fig.~\ref{fig:t1-reboot} shows one of them. The
red line shows the percentage of packets the switch dropped, as seen
by Clustermapper. The switch stopped forwarding packets, though it
remained up.  This was detected by automation and it was rebooted at
t=0.  The switch fully resumed forwarding by t=2 mins. The purple line
shows the percentage of packets dropped each minute, multiplied by
10K to make it visible. Around a quarter of QPs were affected and
around 580K packets were dropped. The blue line shows total job
throughput: it dips when the switch initially failed, as many QPs lost
packets, but they quickly mapped out the bad path and throughput was
largely unaffected afterwards. When the switch actually rebooted,
there was no impact.

Despite this resiliency, some single points of failure remain. We use
800 Gb/s NIC optics, split into 4x200 Gb/s links. If the NIC
transceiver itself flaps, we lose all ports on the NIC, and cannot
ride this out because QPs fail. We do see such events when training at
scale, but fortunately they are rare.
A different optical design might avoid this issue altogether.

%% file: experiments_nvidia.tex
\subsubsection{Point-to-Point Communication Performance}
\label{sec:performance}

The table below presents the point-to-point latency and
bandwidth communication performance of MRC on CX-8, measured on Cluster~B using
the \texttt{ib\_write\_lat} and \texttt{ib\_write\_bw} perftest between a
single client and a single server with one and four QPs, respectively. \vspace{0.1in}

\begin{tabular}{l l l l}
	\hline
	\textbf{Topology} & \textbf{Message Size} & \textbf{Metric} & \textbf{Result} \\
	\hline
	T0-Local  & 2\,B  & Latency   & 5.09 \,\textmu s \\
	T0-Local  & 32\,KB  & Bandwidth & $\approx$ 770\,Gb/s \\
	
	Cross-T1  & 2\,B  & Latency   & 6.54 \,\textmu s \\
	Cross-T1  & 32\,KB  & Bandwidth & $\approx$ 770\,Gb/s \\
	\hline
\end{tabular}

\vspace{0.1in}The bandwidth test performs back-to-back writes of a fixed 32\,KB message
size and reports the achieved application-level GPU-to-GPU bandwidth. Both
{\em T0-local} and {\em cross-T1} configurations achieve approximately
770\,Gb/s, corresponding to 96\% of the theoretical peak bandwidth. This
indicates that steady-state throughput is not constrained by whether traffic
remains within a single T0 or traverses T0 boundaries.

The latency test uses two-byte messages and measures the time from posting the
write until completion is delivered at the sender, dividing the result by two
to approximate one-way latency. {\em T0-local} communication achieves a latency
of 5.09\,$\mu$s, while {\em cross-T1} communication exhibits a higher latency
of 6.54\,$\mu$s. The T0-local latency reflects fixed MRC control-path overheads such
as queue pair processing, work request handling, and path management. The
additional latency observed in the cross-T1 case is attributable to extra
switch hops when crossing between T0 domains. These effects modestly impact
short-message latency while having negligible effect on large-message
bandwidth.

\begin{figure*}
\hspace{0.05\textwidth}
  \begin{subfigure}{0.45\textwidth}
	\includegraphics[width=0.95\textwidth]{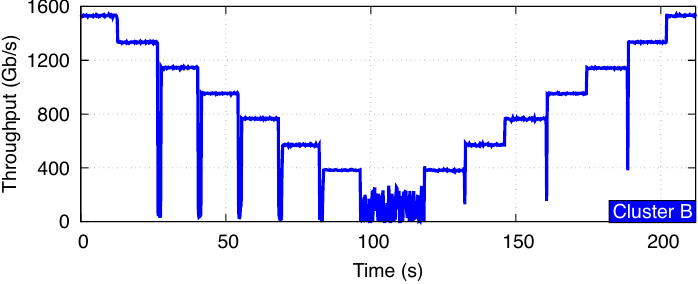}
	\captionsetup{font=scriptsize}
	\caption{T0-local: NIC–T0 link fail/recovery}
	\label{fig:nic-t0-linkdown}
  \end{subfigure}
  \begin{subfigure}{0.45\textwidth}
	\includegraphics[width=0.95\textwidth]{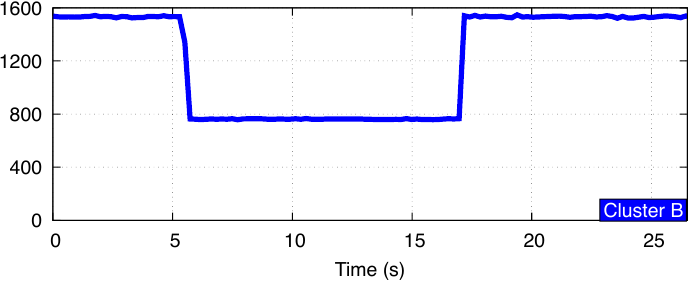}
	\captionsetup{font=scriptsize}
	\caption{T0-Local: Four NIC Links Flap}
	\label{fig:nic-t0-flap}
  \end{subfigure}

\medskip

\hspace{0.05\textwidth}
  \begin{subfigure}{0.45\textwidth}
	\includegraphics[width=0.95\textwidth]{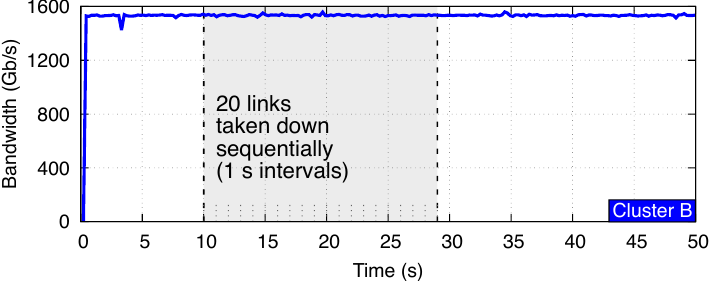}
	\captionsetup{font=scriptsize}
	\caption{Cross-T1: T0–T1 link fail}
	\label{fig:t0-t1-linkdown}
  \end{subfigure}
  \begin{subfigure}{0.45\textwidth}
	\includegraphics[width=0.95\textwidth]{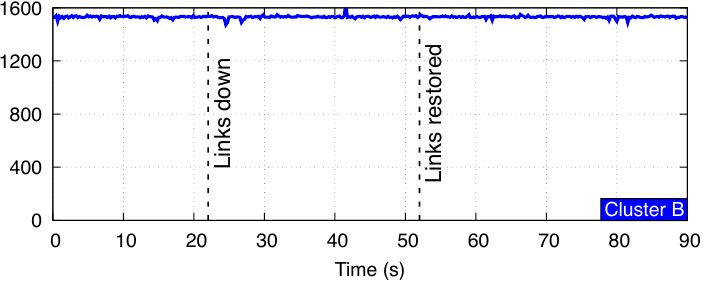}
	\captionsetup{font=scriptsize}
	\caption{Cross-T1: T0–T1 link flap}
	\label{fig:t0-t1-flap}
  \end{subfigure}
	\caption{T0-Local and Cross-T1 Reliability Results with ib\_write\_bw (bi-directional)}
\end{figure*}

\subsubsection{MRC Response to Link Down and Flap Events}
\label{sec:reliability_1}

We study how MRC responds to network failures on Cluster~B, including link
down and link flap events at different points in the network, as well as
transceiver flaps. Results for similar experiments run on Cluster~D are shown
in the Appendix. 

We first examine how the throughput of a transfer using four QPs
degrades gracefully as NIC--T0 links fail. We run bidirectional
\texttt{ib\_write\_bw} between two T0-local NICs and sequentially bring
NIC--T0 links down. As shown in Fig.~\ref{fig:nic-t0-linkdown},
throughput decreases as links are taken offline one at a time and
recovers as they are restored. On link failure, MRC detects the link
down event and re-balances EVs across the remaining
planes. It also updates the link-state bitmap in SACK packets to notify
the remote endpoint that the link is unusable, prompting the remote MRC
instance to remap its EV set to avoid the failed plane. We observe a
brief interruption of less than a second during failure detection, EV
remapping, and packet retransmission, after which both inbound and
outbound QPs stabilize over the remaining links. As links recover,
traffic in both directions quickly resumes using them.

In Fig.~\ref{fig:nic-t0-flap}, we flap four of the eight links on a single
CX8 NIC to emulate a realistic production failure scenario in which multiple
links may be impacted simultaneously due to a shared OSFP port or transceiver.
Such failures can cause all links associated with the port to go down at once,
resulting in the loss of up to four links concurrently. 
As shown in Fig.~\ref{fig:nic-t0-flap}, MRC quickly stabilizes at
approximately half the nominal bandwidth when all four links are taken down
(around the 5 second mark), and rapidly recovers once the links are restored
(around the 17 second mark). While link flap durations in practice are typically
much shorter, this experiment synthetically induces a flap by disabling the
links and re-enabling them after a fixed interval.

Fig.~\ref{fig:t0-t1-linkdown} illustrates the impact of T0–T1 link failures
on MRC traffic. Because MRC sprays packets from each QP across a large number
of paths, T0–T1 link failures have a substantially smaller impact than NIC–T0
failures.  In this experiment, we run cross-T1 \texttt{ib\_write\_bw} and
synthetically disable twenty links sequentially. Telemetry collected during the
run indicates that the majority of these links were actively carrying traffic
prior to failure, demonstrating broad utilization of the available paths. As
links are taken down, MRC maps out the affected EVs and retransmits lost
packets over alternate paths.  Bandwidth is reported at approximately 200\,ms
granularity, and we observe minimal impact on overall throughput even as active links
are removed.

Fig.~\ref{fig:t0-t1-flap} also shows the effect of flapping eight T0–T1 links
simultaneously and restoring them after a few seconds, as may occur due to a
switch transceiver flap. Telemetry shows that these links were carrying
traffic prior to the flap.  As in the previous experiment, MRC reroutes traffic
and recovers rapidly, resulting in negligible impact on workload-level
performance.

\subsubsection{MRC Behavior with T0/T1 Switch Failures}

Although uncommon, switches may become unresponsive or crash during
production workloads in large-scale training clusters. To evaluate the
resulting impact on workload performance, we conducted experiments on
Cluster~B in which T0 and T1 switches were taken down while running
\texttt{ib\_write\_bw} to drive network traffic.

\begin{figure}[t]
  \centering
  \includegraphics[width=0.9\linewidth]{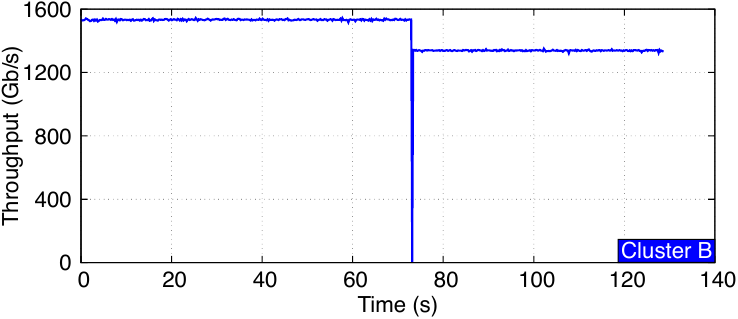}
	\caption{T0-Local ib\_write\_bw during T0 switch failure.
	}
  \label{fig:t0-switch-down}
\end{figure}

Fig.~\ref{fig:t0-switch-down} illustrates the impact of taking a T0 switch
down during the experiment. After EV remapping completes, the steady-state
bandwidth drops by approximately 100\,Gb/s, reflecting the removal of EVs
associated with the failed T0 switch from the active EV set. Throughput then
stabilizes at a level proportional to the remaining available network
capacity. This behavior closely mirrors the results presented in
Fig.~\ref{fig:nic-t0-linkdown}, confirming that MRC failure handling is
driven by end-to-end path availability rather than the specific location or
type of underlying fabric fault. As a result, switch-level failures manifest
as a predictable reduction in usable path diversity, while preserving
application progress and maintaining stable throughput over the remaining
healthy paths.

\begin{figure}[t]
  \centering
  \includegraphics[width=0.9\linewidth]{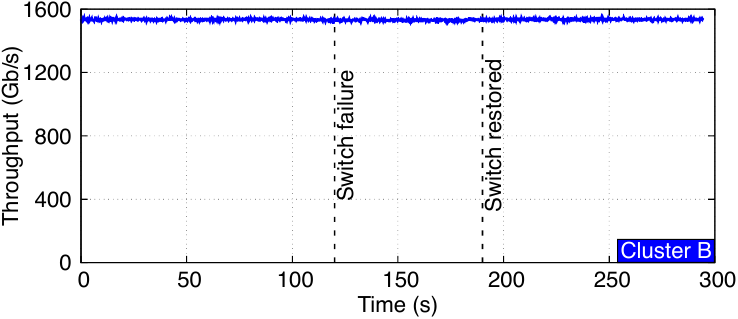}
  \caption{Cross-T1 ib\_write\_bw during T1 failure/recovery.
	}
  \label{fig:t1-switch-down}
\end{figure}

Fig.~\ref{fig:t1-switch-down} illustrates the impact of taking a T1 switch
down during a cross-T1 \texttt{ib\_write\_bw} experiment with four QPs. 
In this scenario, each QP is sprayed
across a large number of paths, and sufficient alternative EVs remain
available to sustain the aggregate network capacity. Consequently, no
steady-state bandwidth degradation is observed despite the failure. Once the
T1 switch is restored, transient fluctuations subside and bandwidth returns
to a stable level.

\subsubsection{Robustness to Path-Level Packet Loss}

To evaluate the robustness of MRC under transient link impairments, we
conducted this experiment on Cluster~B using NVIDIA’s MRC debug capability to
inject controlled errors at the EV level. Specifically, we configured a
selected EV to drop 20\% of its packets and measured the resulting EV state
transitions and end-to-end performance using the cross-T1 bidirectional
\texttt{ib\_write\_bw} benchmark. To isolate and clearly observe EV status
updates, we restricted the system to a small fixed set of 16 paths,
forcing just two paths per each of the eight planes.

\begin{figure}[t]
  \centering
  \includegraphics[width=0.9\linewidth]{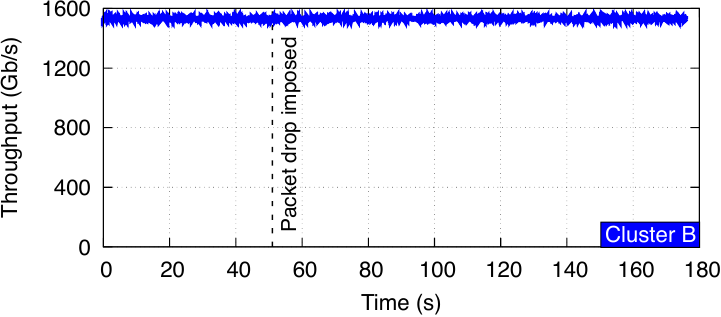}
  \caption{Packet-drop reliability experiment.
	}
  \label{fig:pktdrop}
\end{figure}

\begin{figure}[t]
  \centering
  \includegraphics[width=0.9\linewidth]{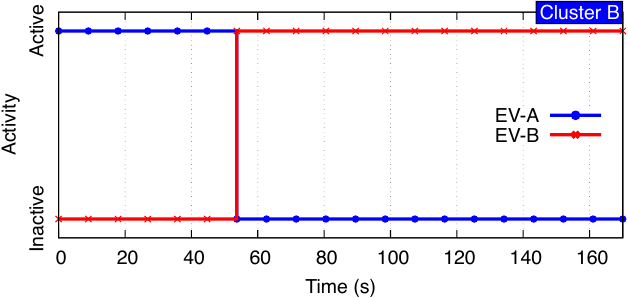}
  \caption{Path activity during packet-drop experiment. 
	}
  \label{fig:pktdrop:state}
\end{figure}

During the experiment, we continuously monitored EV status to capture how the
system reacts to injected faults. As shown in Fig.~\ref{fig:pktdrop:state},
EV-A is initially active while EV-B remains inactive. At approximately 51~s,
when packet drop is induced on EV-A, its status transitions immediately to
inactive, and EV-B is activated as a replacement. This behavior demonstrates
MRC’s ability to promptly detect faults and switch traffic to an alternative
EV, aligning with the sustained line-rate bandwidth observed in
Fig.~\ref{fig:pktdrop}.

We repeated the experiment with different packet drop rates and observed the
same qualitative behavior: the affected EV was promptly removed from the active
set, traffic was transparently redirected to alternative paths, and
application-level bandwidth remained stable.

\subsubsection{Load Balancing Across EVs}
\label{sec:exp:loadbalance}

We evaluate MRC's ability to dynamically balance load across EVs using a
controlled two-flow experiment in Cluster B. The setup consists of two
communication pairs operating over two EVs (EV-A and EV-B) within the same
plane.  We first establish an initial flow between \texttt{client1} and
\texttt{server1}. In the steady state, this flow is mapped entirely onto EV-A,
achieving near line-rate bandwidth while EV-B remains idle. At approximately 65~s, 
we introduce a second flow between \texttt{client2} and
\texttt{server2}, which is explicitly forced to use EV-A. This creates
transient congestion, with both flows contending for the same EV.

\begin{figure}[t]
  \centering
  \includegraphics[width=0.9\linewidth]{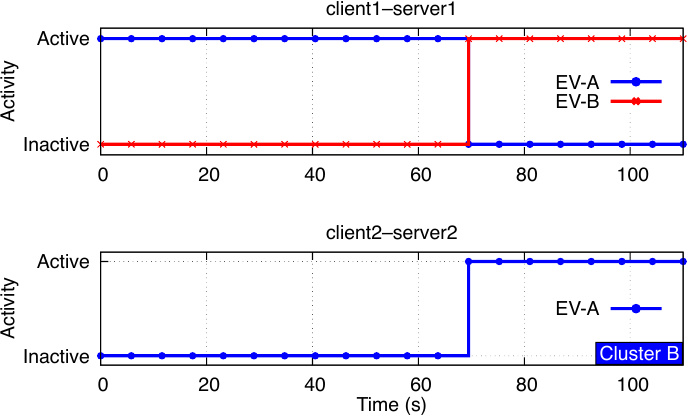}
  \caption{Path activity during load balancing experiment. 
	}
  \label{fig:loadbalance:evstate}
\end{figure}

Upon detecting this congestion via ECN, MRC triggers traffic redistribution to rebalance
load across the available EVs. Specifically, the
\texttt{client1}-\texttt{server1} flow is migrated to EV-B, while the
\texttt{client2}-\texttt{server2} flow continues on EV-A. This transition is
clearly visible in the per-EV activity traces shown in
Fig.~\ref{fig:loadbalance:evstate}: EV~A transitions from serving a single
flow to hosting the second flow exclusively, while EV-B becomes active as it
assumes responsibility for the migrated flow.  Importantly, this redistribution
occurs without observable performance degradation.  Both client-server flows
sustain near-peak bandwidth throughout the transition, with aggregate
throughput remaining close to line rate. The results demonstrate that MRC can
effectively rebalance traffic across EVs in response to dynamic load changes,
maintaining stable, high throughput with minimal disruption.

\subsubsection{NCCL Collective Execution at Scale}
\label{sec:exp:nccl}

\begin{figure}[t]
  \centering
  \includegraphics[width=0.9\linewidth]{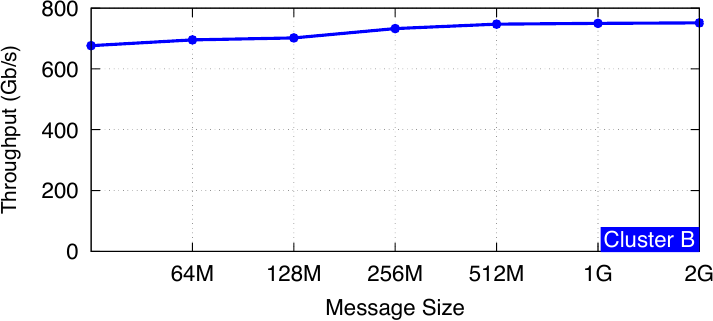}
	\caption{NCCL send/recv performance at 42K GPUs scale.
	}
  \label{fig:exp:ncclsendrecv}
\end{figure}

We also conducted NCCL microbenchmark experiments on Cluster~B to evaluate
the scalability of MRC. Specifically, we used the NCCL-tests \texttt{sendrecv}
benchmark, in which each node concurrently sends data to and receives data
from a neighboring peer, forming a steady-state bidirectional communication
pattern. The reported bandwidth corresponds to the measured per-NIC throughput.
As shown in
Fig.~\ref{fig:exp:ncclsendrecv}, NCCL over MRC achieves up to 92\,GBytes/s for
large message sizes at a scale of 42K GPUs.

%% file: experiments-amd.tex
\subsubsection{Comparison with RoCE}

\begin{figure}[t]
  \centering
  \includegraphics[width=0.9\linewidth]{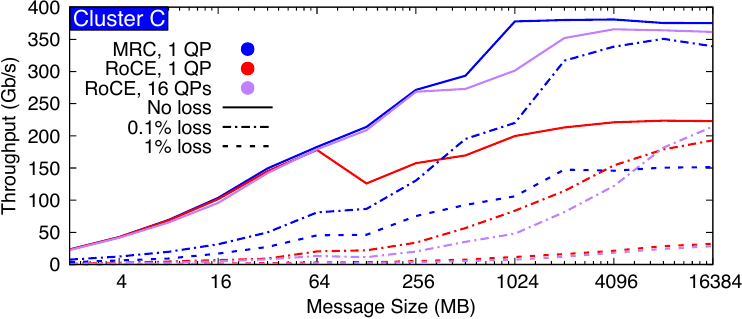}
  \caption{MRC and RoCE performing 64-way ring all-reduce, for varying message size and loss rates}
  \label{fig:roce-ar}
\end{figure}

\begin{figure}[t]
  \centering
  \includegraphics[width=0.9\linewidth]{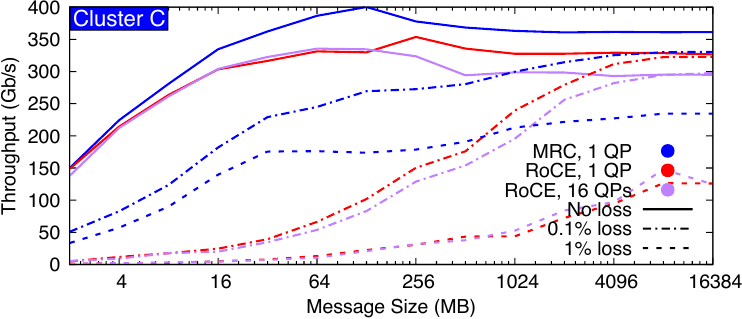}
  \caption{MRC and RoCE performing 64-way all-to-all, for varying message size and loss rates}
  \label{fig:roce-a2a}
\end{figure}

\begin{figure*}
\begin{subfigure}{0.33\textwidth}
  \includegraphics[width=\textwidth]{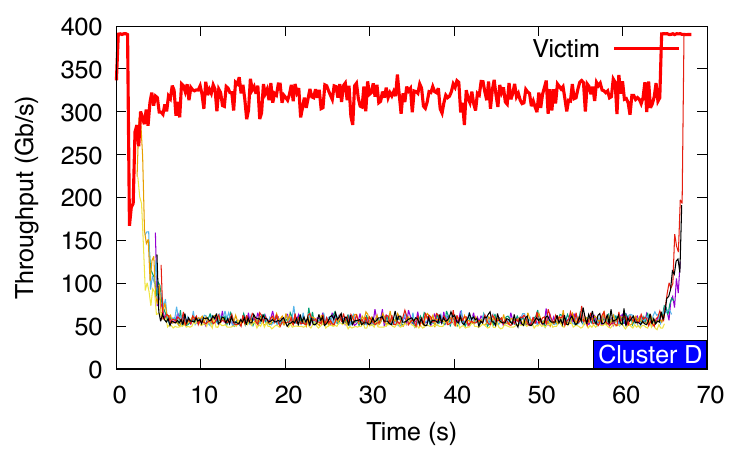}
  \caption{RoCEv2 with DCQCN, 1 QP}
  \label{fig:incastdcqcn1qp}
\end{subfigure}
\begin{subfigure}{0.33\textwidth}
  \includegraphics[width=\textwidth]{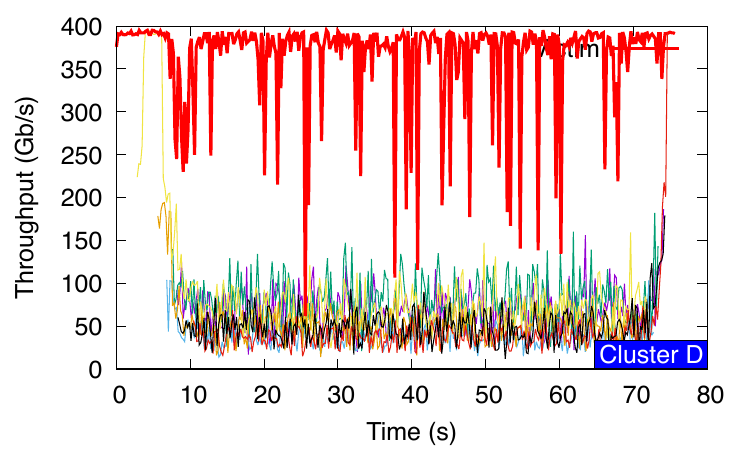}
  \caption{RoCEv2 with DCQCN, 10 QPs}
  \label{fig:incastdcqcn8qp}
\end{subfigure}
\begin{subfigure}{0.33\textwidth}
  \includegraphics[width=\textwidth]{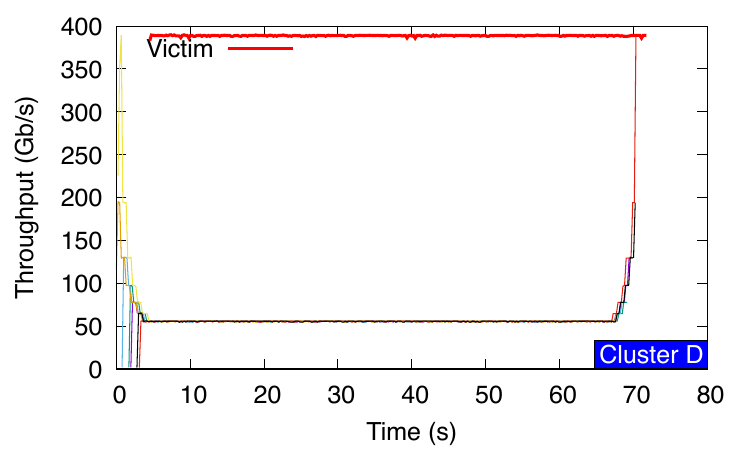}
  \caption{MRC, 1 QP}
  \label{fig:incastmrc1qp}
\end{subfigure}
\caption{7 to 1 incast with a victim flow destined to a different node in the same rack.}
\end{figure*}

We don't have a large deployment that can perform a direct comparison
between MRC and RoCE, but we built two small scale testbeds, one using AMD
Pollara NICs (Cluster C) and one using Broadcom Thor Ultra NICs (Cluster D), to compare.

Cluster C consists of 64 GPUs, each paired with a Pollara 400 Gb/s
NIC, using TH5 switches in a two-tier Clos topology. For RoCE, we
configure a single-plane network, as is common, with 400 Gb/s links
and ECMP routing.  In this configuration, PFC is enabled and DCQCN is
used as congestion control to avoid excessive PFC activation. For MRC,
we configure a four-plane network with four 100 Gb/s links per NIC; we
disable PFC and use SRv6 routing. This is the closest like-for-like
comparison we can build while giving each protocol its preferred
environment: the GPUs, NICs, switches, and aggregate bandwidth are
identical.

We wish to understand two things. First, does MRC load balance better
than RoCE, even with QP scaling? Second, does MRC's selective
retransmit really help typical AI collectives in the presence of
network problems?  To answer this, we run two sets of
experiments. First, we run all-reduce on a ring, which generally
requires each node to send to one node and receive from one node
simultaneously, with the goal that all transfers reach line rate. Any
glitch will cause a bubble and hurt overall performance. Second, we
run all-to-all, which requires every node to send to and receive from
multiple nodes simultaneously. The goal is to observe load balancing
and retransmit performance during a collective that can cause incast
congestion.

Fig.~\ref{fig:roce-ar} shows mean bandwidth when performing
all-reduce for increasing message sizes.  For MRC we use one QP,
whereas for RoCEv2 we show the results using one QP and 16 QPs.  The
solid curve shows the baseline results with no induced packet loss.
For small message sizes, the collective is latency-bound, and there is
little difference between RoCE and MRC.  For larger message sizes we
become increasing bandwidth-bound.  RoCE with one QP suffers from ECMP
hash collisions and generally achieves only half the possible
throughput.  The conventional way to improve load balancing with RoCE
is by QP scaling - load balancing a transfer across multiple QPs to
reduce the impact of flow collisions.  We find that, for RoCE, QP
scaling does indeed help, but we see very little gain beyond 8
QPs (the figure shows 16).  In contrast, one MRC QP spraying across 256 paths
achieves better performance than 16 QPs, primarily due to doing a good
job of load-balancing the network.

The dashed lines show performance as we add packet loss at 0.1\% and
1\% levels.  We do this by adding an inline P4\cite{p4} program to all NICs in the cluster
that randomly drops incoming packets at a prescribed rate.  RoCE
was not designed to be resilient to packet loss and, as expected, it
suffers poor performance.  MRC is more resilient - with large message
sizes it can retransmit fast enough that 0.1\% loss has little impact.
At smaller message sizes the collective is more latency-bound, and
recovering packet losses has more impact due to tail losses not
being possible to mask.  MRC still performs relatively well though.
We don't expect an AI cluster to have persistent high loss rates, but we do see a range of causes of loss transients; synchronous pretraining does not care about mean performance---only the tail matters---so at scale, having a protocol that tolerates loss is important.

At 1\% loss RoCE is pretty much ususable and even MRC only gets around
a third of the intended throughput.  This is enough to ride out a
short transient burst of loss, but not good enough for continued
training.  It is worth noting that this test shows worse performance
than MRC would get in typical real-world scenarios because the loss is
applied on all planes simultaneously.  If a T0-T1 link is dropping at
a high rate, MRC will simply stop using it, as in Fig.~\ref{fig:pktdrop:state}.  If a NIC-T0 link is
dropping at a high rate, our Clustermapper monitor will detect it, and a denylist
entry can be added to avoid the broken plane, limiting the performance
degradtion to just the fraction of bandwidth lost by dropping one
plane.  Whether this is sufficient to allow the node to persist in a
training job is then a policy decision.

Fig.~\ref{fig:roce-a2a} shows mean bandwidth when performing
all-to-all for increasing message sizes.  Again the solid lines are
baseline with no loss, and the dashed lines add random loss.  In this
case more QPs are active simultaneously, so RoCE's load balancing
is not such an issue.  Indeed, we see that QP scaling is not helping
RoCE: 16 RoCE QPs per destination actually perform slightly worse than
one QP and performance starts to drop off slightly with more than two
QPs.  For all message sizes MRC outperforms RoCE, but the difference
is greater in the regime that is bandwidth-bound.  

With loss, RoCE suffers worse than MRC, as expected, but the
difference is much greater at smaller message sizes.  At large message
sizes and 0.1\% loss, RoCE barely sees any degradation.  This is
because so many QPs are active simultaneously that few packets are in
flight on each one. When the transfers are large enough one QP pausing for
retransmission is mostly masked by others being active.  At small
message sizes, there isn't time to mask loss, and RoCE does very
poorly.  MRC's SACK-based retransmission helps greatly here, despite
this test causing loss on all planes.

%% file: experiments-collateral.tex
\subsubsection{Collateral Damage}
\label{sec:collateral}

Multiple collectives performing different axes of training parallelism may run
simultaneously and can interfere with each other.  This can particularly be a problem with PFC, which
MRC was designed to avoid.  Lossless networks struggle with incast traffic patterns when the
congestion spreads and can affect unrelated ``victim'' traffic. To
test this behaviour we run a cross-spine 7 to 1 incast traffic
pattern, and in parallel we have another ``victim'' connection to an
idle destination in the same rack as the incast target.

The testbed for this experiment is Cluster 4, a small-scale testbed
with 16 servers each with one RTX6000 GPU and a Broadcom Thor Ultra NIC. We
use VRFs on Broadcom TH5 switches to emulate a single-plane 2-tier
Clos topology with 4 racks each with 4 servers and four spine
switches.  Due to server limitations we run all links in this testbed
at 400 Gb/s.

With RoCE, if we do not use DCQCN and rely solely on PFC to manage
congestion, the victim flow is pulled down to a rate not much greater
than the incast flows (see Appendix for details). DCQCN is designed to
greatly reduce PFC, and we find it does help, but is not sufficient. With a
single QP (Fig.~\ref{fig:incastdcqcn1qp}), victim-flow performance still
degrades by about 25\%. With eight QPs
(Fig.~\ref{fig:incastdcqcn8qp}), sharing is worse, but the average
impact on the victim flow is smaller; however, there are one-second
intervals where the victim's throughput is 100Gbps, a 75\% drop from
optimal. In both cases, DCQCN cannot properly control the bottleneck
queue and still generates some PFCs. In contrast, MRC
(Fig.~\ref{fig:incastmrc1qp}) almost perfectly shares the bottleneck link
among the incast flows and has no impact on the victim flow.

In principle DCQCN parameter tuning should reduce this issue. However,
configuring DCQCN properly is very hard because it is traffic pattern
specific, to the point that some hyperscalers have disabled it in
production \cite{meta-roce}. We provide some tuning results in the Appendix
that illustrate this.

%% file: relatedwork.tex
\section{Related Work}

Load balancing effectiveness is defined by distribution granularity:
per-flow, per subflow and per packet. Single path transports (RoCEv2)
using per-flow ECMP \cite{ecmp} are simple but
collision-prone. Solutions to improve flow placement span centralized
(Hedera \cite{hedera} or MicroTE \cite{microTE}), switch-based
(Flowlet switching \cite{201562}, Presto \cite{flowcell} and
DLB\cite{dlb}) and host-based approaches (e.g. PLB
\cite{10.1145/3544216.3544226}, Flowcut~\cite{11299491}, and
FlowBender \cite{flowbender}). All these approaches often react too
slowly for bursty AI traffic, and have had limited success under high
load. Multipath operation is needed for collisions to be efficiently avoided.
Switch-based multipathing solutions exist that provide in-order per flow delivery to hosts
(Drill \cite{drill}, CONGA \cite{conga} and Stardust \cite{stardust};
commercial offerings from Broadcom and Cisco) - however, these are
proprietary solutions and require homogeneous networks.

Many deployments today rely on RoCEv2 with ECMP routing but use
application-level multipath transports, typically implemented in the collective
communication layer to reduce the effects of colisions, e.g.
NCCL~\cite{nccl} QP scaling, MSCCL \cite{msccl}, NCCLX \cite{ncclx},
UCCL \cite{uccl2}. Application-level multipathing mitigates the issues but does
not completely solve them, as we have shown in our evaluation.

Multipath TCP (MPTCP~\cite{mptcp-dc}) improves utilization by spreading
a single connection across multiple subflows.  MPTCP and similar
approaches must keep state per subflow however, but they can be used
for any datacenter topology and work particularly well with long
running traffic.  Google’s Falcon~\cite{falcon} uses this approach to
implement a multipath replacement for RoCEv2 in the NIC.

Per-packet load balancing (or packet spraying) has been proposed
for Clos networks to reduce the amount of state. RPS \cite{spray}, Homa
\cite{homa} and NDP \cite{ndp} are oblivious to path state,
but struggle with asymmetric congestion and partial/gray failures.  By
keeping a small amount of state per (virtual) path, feedback-driven methods like MPRDMA \cite{mp-rdma},
Hermes \cite{hermes}, REPS\cite{reps} or Strack utilize ECN or delay
signals to move traffic away from congested or failed paths.

MRC changes RoCEv2 to enable packet-level load balancing, selective
retransmissions and improve reliability in multi-plane networks; its
target deployment is best-effort (i.e. lossy) networks that support
packet trimming. To achieve this, MRC builds upon a wide body of prior
work as follows.  IRN was one of the first approaches to add selective
retransmission to RoCEv2 \cite{irn}, while MPRDMA added multipath
operation and selective retransmission as well as keeping per path
state \cite{mp-rdma}.
These approaches still rely on lossless networks,
but it is difficult to completely avoid HoL blocking issues.

MRC is similar to the Ultra Ethernet Transport, an industry-driven
standard that replaces RoCEv2 with a brand new protocol stack aiming
to support both HPC and AI workloads \cite{uet}. UET has standardized
packet trimming, and also targets operation in best-effort networks.

In contrast to most existing works which use host-driven ECMP routing or switch
spraying, MRC uses source routing with SRv6 to increase robustness at scale.  Filsfils et
al.\ \cite{filsfils_srv6_ai} have validated SRv6 micro-segment (uSID) based path placement for single path RoCEv2.

We are not the first to propose topology co-design for AI network (see \cite{99probs} for an
overview of this space). Alibaba's HPN\cite{hpn} uses a dual-ToR, rail optimized
networks, connecting up to 15K GPUs in a two-tier design. Wang et al.\cite{railonly} propose using a rail-only
approach to have a single tier of switches. However, we are among the first to
design and deploy a multi-plane network to reach 100K+ GPUs with two switch tiers.

Experiences from multiple hyperscalers have highlighted the negative
effect of failures on training jobs \cite{llama3,hpn}.  Our
multi-plane, static-routed SRv6 approach directly targets being able
to ride out network failures.

%% file: conclusions.tex
\section{Conclusions}

MRC is designed to load balance a multi-plane network by spraying each
QP across all planes and many paths in each plane, performing
fine-grain active load balancing and routing around failures. We
implemented MRC in 800Gb/s NICs from Nvidia, Broadcom, and AMD, and
built multiple supercomputers that use a two-tier multi-plane topology
running MRC in the back-end network. MRC's ability to route around
failures allowed us to disable dynamic routing; instead, we use SRv6
source routing with static routes in the switches. These
supercomputers have been used to train OpenAI's latest frontier models, and
we have observed that this design allows very large AI pretraining
jobs to ride out network failures that would previously have caused
the job to fail, while most failures have minimal impact on job step
time. We find that static source routing gives us very good
observability and reduces operational burden, while MRC's resilience
means that many network failures are not even urgent to repair.

%% file: acknowledgments.tex
\section{Acknowledgments}

Developing and successfully deploying MRC has involved a great many people;  we would particularly like to thank the following (in alphabetical order): 
\input{ack-people-sorted}

%% file: ack-people-sorted.tex
Rukhsana Ansari,
Dragos Argint,
Cristi Baciu,
Bar Becker,
Omri Ben~David,
Shai Ben~Haim,
Paul Blakey,
Jeremias Blendin,
Brian Box,
Greg Brockman,
Mihai Brodschi,
Evan Burness,
Trevor Cai,
Rory Carmichael,
Miguel Castro,
Peng Cheng,
James Crooks,
Janet Cui,
Valerie Cutts,
Michael Dalton,
Biswa Dash,
Shawn Dashuai Zhang,
Mark Debbage,
Karl Deng,
Weixin Deng,
Gregor Dick,
Saurabh Dighe,
Qixin Dong,
Gili Doweck,
Iulian Dracea,
Peter Dunning,
Yakov Dyadkin,
Madan Easwaramoorthy,
Elliot Edmunds,
Sally Egan ,
Lior Erets Kdosha,
Ze Gan,
Naren Gathoo,
Renaud Gaubert,
Ahmad Ghalayini,
Jeff Glover,
Guru Harakere,
Damian Hazen,
John Huber,
Richard Hughes,
Rita Hui,
Tony Hurson,
Changho Hwang,
Iva Ivanov,
Vivek Jain,
Riff Jiang,
Anuj Kalia,
Ali Kamali,
Nemanja Kamenica,
Vivek Kashyap,
Bhunu Kathavarayan,
Ady Khalifa,
Xinhao Kong,
Shilpa Kothapalli,
Gawaskar Kumar,
Ariel Levkovich,
Binyang Li,
Xin Liu,
Jie Mao,
Ilias Marinos,
Charlie Mbariky,
Scott McDaniel,
John Mead,
Sharad Mehrotra,
Luke Melton,
Yan Mo,
Scott Moe,
Malek Musleh,
Nikhil Nanal,
Suresh Nedunchezhian,
Duc~Phong Nguyen,
Lisa Nguyen,
Wael Noureddine,
Vlad Olteanu,
Shane O’Neil,
Shahar Oren,
Kumaresh Perumal,
Jonas Pfefferle,
Rajesh Pukhraj Jain,
Catalin Puscoci,
Melur Raghuraman,
Mahdi Ramezani,
David Riddoch,
Uday Ruddarraju,
Kathryn Russell,
Neelabh Sahay,
Lorenzo Saino,
Rafael Salas,
Siva Santosh Pyla,
Emnaual Scaria,
Brent Schartung,
Karen Schramm,
Hemal Shah,
Gilad Shainer,
Sanjay Shanbhogue,
Rohit Sharma,
Eden Shimoni
Delna Sholapurwalla,
Pradeep Sindhu,
Prince Sunny,
Vijay Swaminathan,
Jason Teplitz,
Gaurav Thareja,
Bejoy Thomas,
Sam Truslow,
Dev Upadhyay,
Vamsi Vadlamuri,
Srihari Vegesna,
Ram Velaga,
Jijun Wang,
Yossi Wortzel,
Changrong Wu,
Weijia Yuan,
Reza Zamani,
Jie Zhang,
Yanzhao Zhang,

%% file: experiments_brcm.tex
\begin{figure*}
  \begin{subfigure}{0.30\textwidth}
  \includegraphics[width=\textwidth]{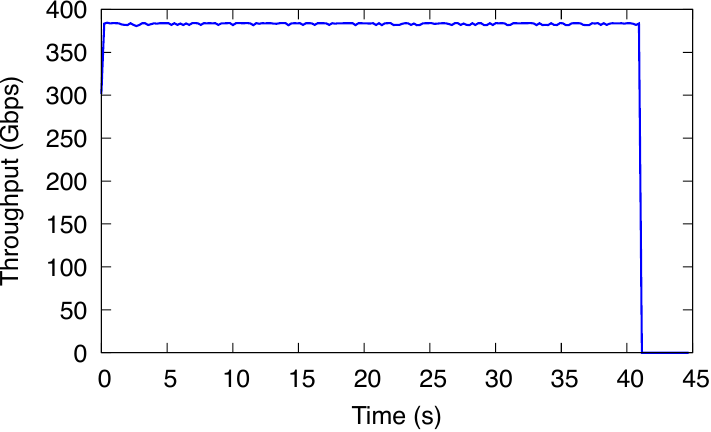}
  \caption{Throughput while sequentially removing three out of four
    400Gb/s T0-T1 links}
  \label{fig:brcm-line}
  \end{subfigure}
  \hspace{0.045\linewidth}
\begin{subfigure}{0.30\textwidth}
  \includegraphics[width=\textwidth]{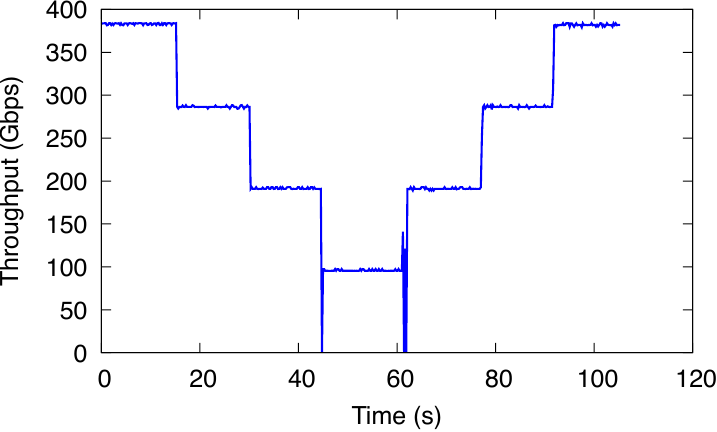}
  \caption{Sequentially removing and then adding three of four 100Gb/s T0-T1 links}
  \label{fig:brcm-steps}
\end{subfigure}
  \hspace{0.045\linewidth}
\begin{subfigure}{0.30\textwidth}
  \includegraphics[width=\textwidth]{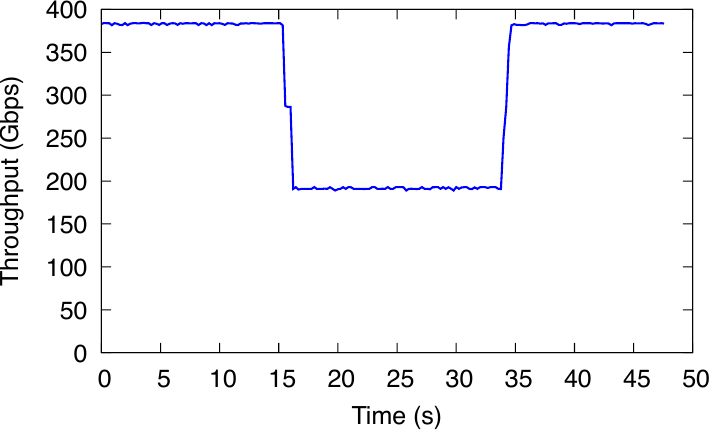}
  \caption{Sequentially removing and then adding three of four 100Gb/s T0-T1 links.}
  \label{fig:brcm-2linksdown}
\end{subfigure}
\caption{ib\_write\_bw performance between two servers in different racks during failures.}
\end{figure*}
\begin{figure*}
\begin{subfigure}{0.33\textwidth}
  \includegraphics[width=\textwidth]{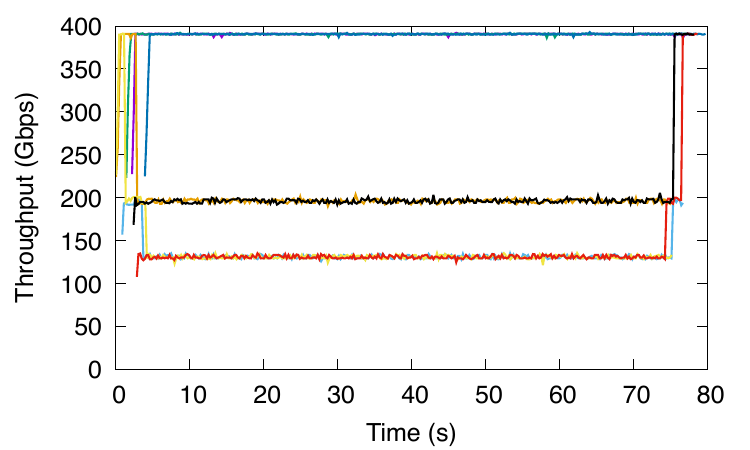}
  \caption{RoCEv2 with PFC, 1QP}
  \label{fig:permpfc1qp}
\end{subfigure}
\begin{subfigure}{0.33\textwidth}
  \includegraphics[width=\textwidth]{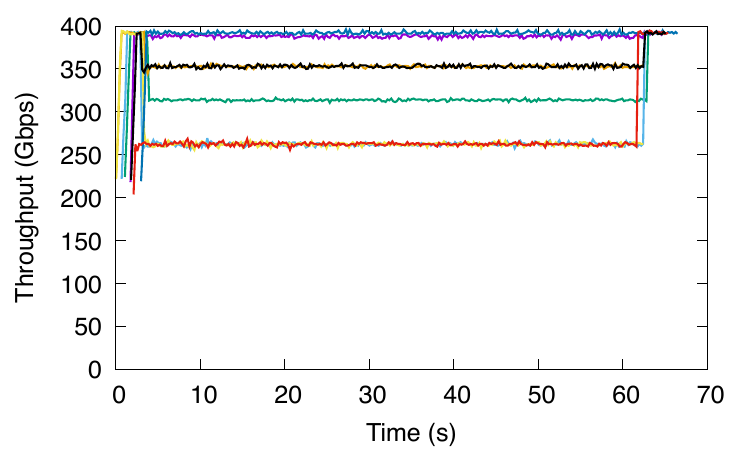}
  \caption{RoCEv2 with PFC, 8QPs}
  \label{fig:permpfc8qp}
\end{subfigure}
\begin{subfigure}{0.33\textwidth}
  \includegraphics[width=\textwidth]{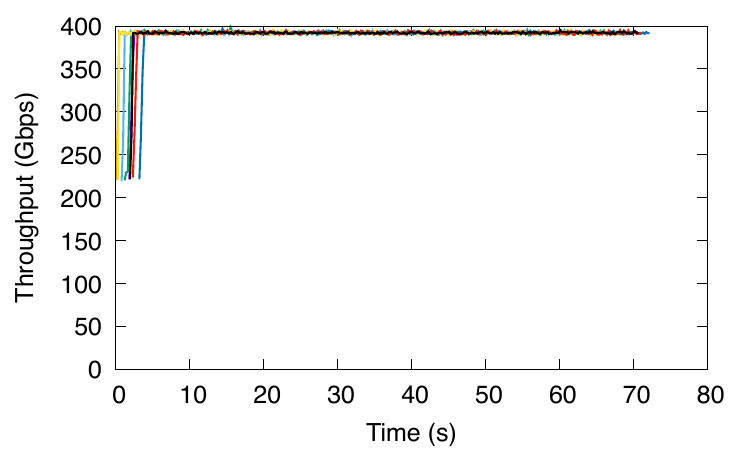}
  \caption{RoCEv2 with DCQCN, 8QPs}  
  \label{fig:permdcqcn8qp}
\end{subfigure}
\caption{Permutation throughput when servers from two racks source flows to servers in other two racks.}
\end{figure*}

We provide some additional results demonstrating MRC's robustness
using Broadcom's Thor Ultra NIC in Cluster D; this is a small 400Gb/s single-plane two-tier network.  MRC uses SRv6, and
we compare with the Thor Ultra plain RoCE implementation running with
PFC and DCQCN enabled.  

We first test the resilience of MRC on Thor Ultra to link failures.
In Figure \ref{fig:brcm-line} NIC--T0 and T0--T1 links are all
400Gb/s, and we show the throughput reported by \textsc{ib\_write\_bw}
as we progressively fail three of the four T0--T1 links. In this case
there is sufficient capacity on the remaining T0--T1 link and, as in
Figure \ref{fig:t0-t1-flap}, MRC on Thor Ultra is able to quickly recover from
failed links with no discernible impact for end-to-end throughput.

In our next experiments we reduce the speed of the T0--T1 links to
just 100 Gb/s, so that as links fail a bandwidth bottleneck is
created.  Figure \ref{fig:brcm-steps} shows the effect of sequentially
disabling three of four links and then adding them back. This
experiment differs from that Figure \ref{fig:nic-t0-linkdown} in that
the failures are not the directly attached NIC links.  MRC is able to
fail onto the remaining path while giving throughput that tracks the
remaining available capacity.  Figure \ref{fig:brcm-2linksdown} shows
a variation of the same experiment where we drop two links at once,
and then bring them back; the results confirm that MRC can track the
available capacity closely.

\begin{figure*}
\begin{subfigure}{0.33\textwidth}
  \includegraphics[width=\textwidth]{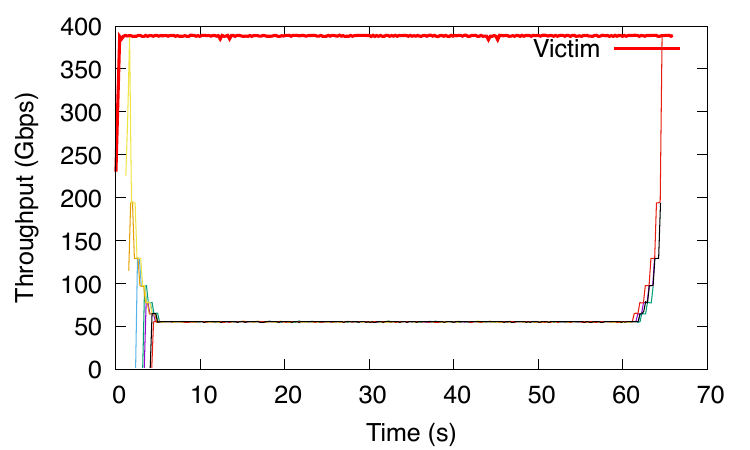}
  \caption{MRC, 8QPs}
  \label{fig:incastmrc8qp}
\end{subfigure}
\begin{subfigure}{0.33\textwidth}
  \includegraphics[width=\textwidth]{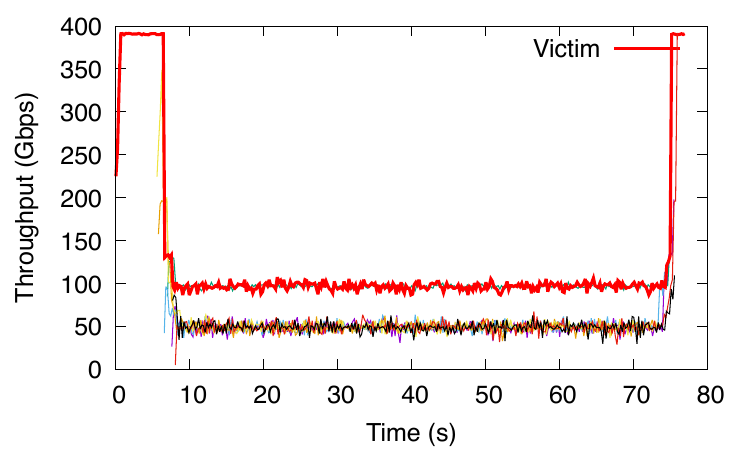}
  \caption{RoCEv2 with PFC only, 1QP}  
  \label{fig:incastpfc1qp}
\end{subfigure}
\begin{subfigure}{0.33\textwidth}
  \includegraphics[width=\textwidth]{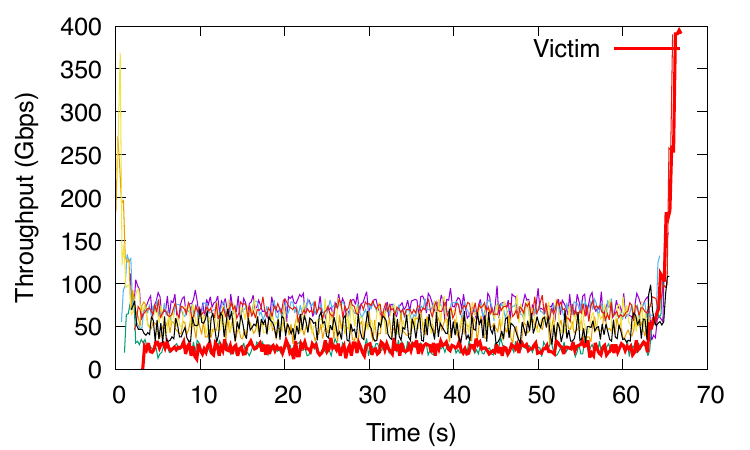}
  \caption{RoCEv2 with PFC only, 8QPs}
  \label{fig:incastpfc8qp}
\end{subfigure}
\caption{7 to 1 incast with a victim flow destined to the same rack.}
\end{figure*}
\begin{figure*}
  \begin{subfigure}{0.33\textwidth}
    \includegraphics[width=\textwidth]{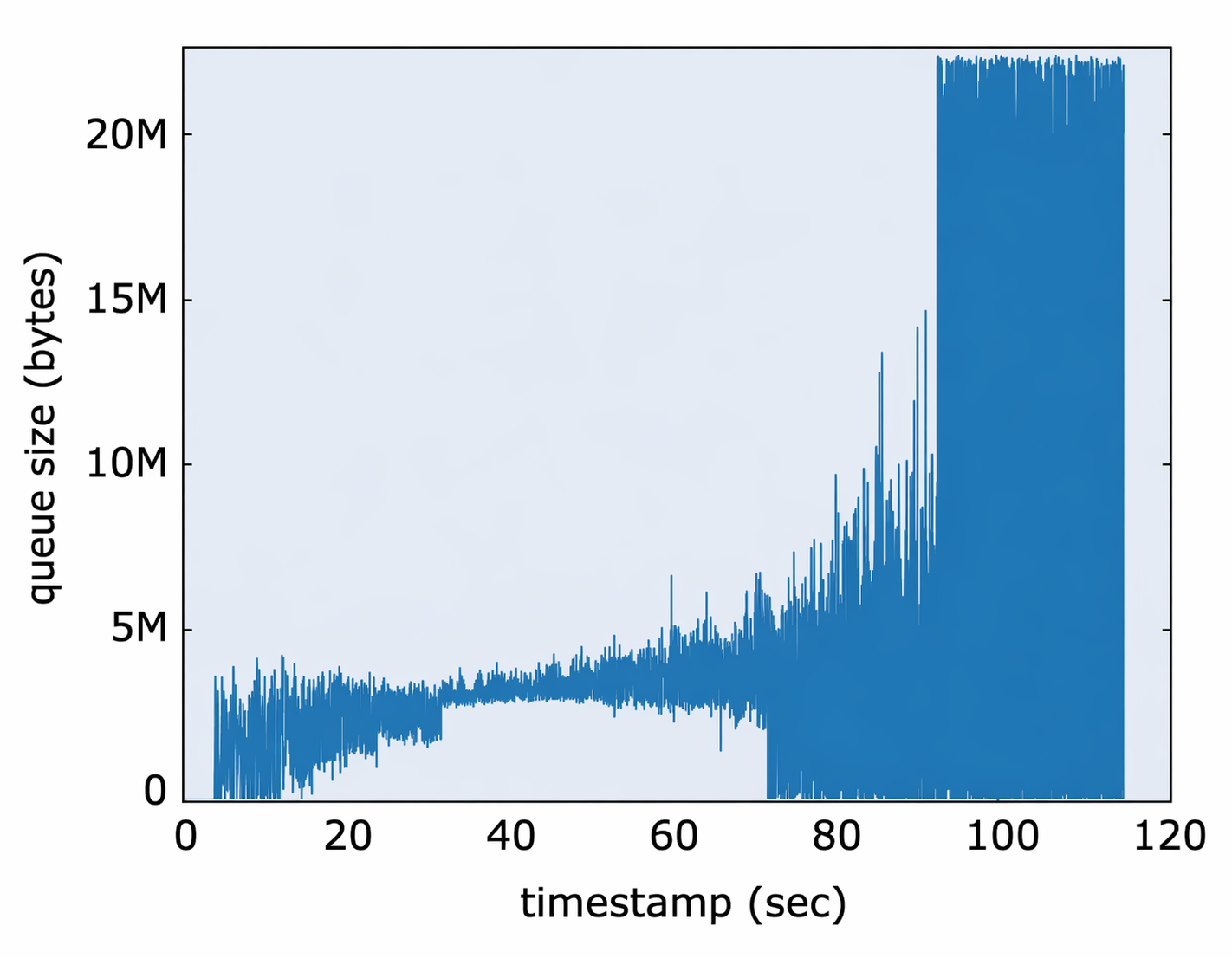}
    \caption{Default DCQCN parameters}
    \label{fig:dcqcn-default}
  \end{subfigure}
  \begin{subfigure}{0.33\textwidth}
    \includegraphics[width=\textwidth]{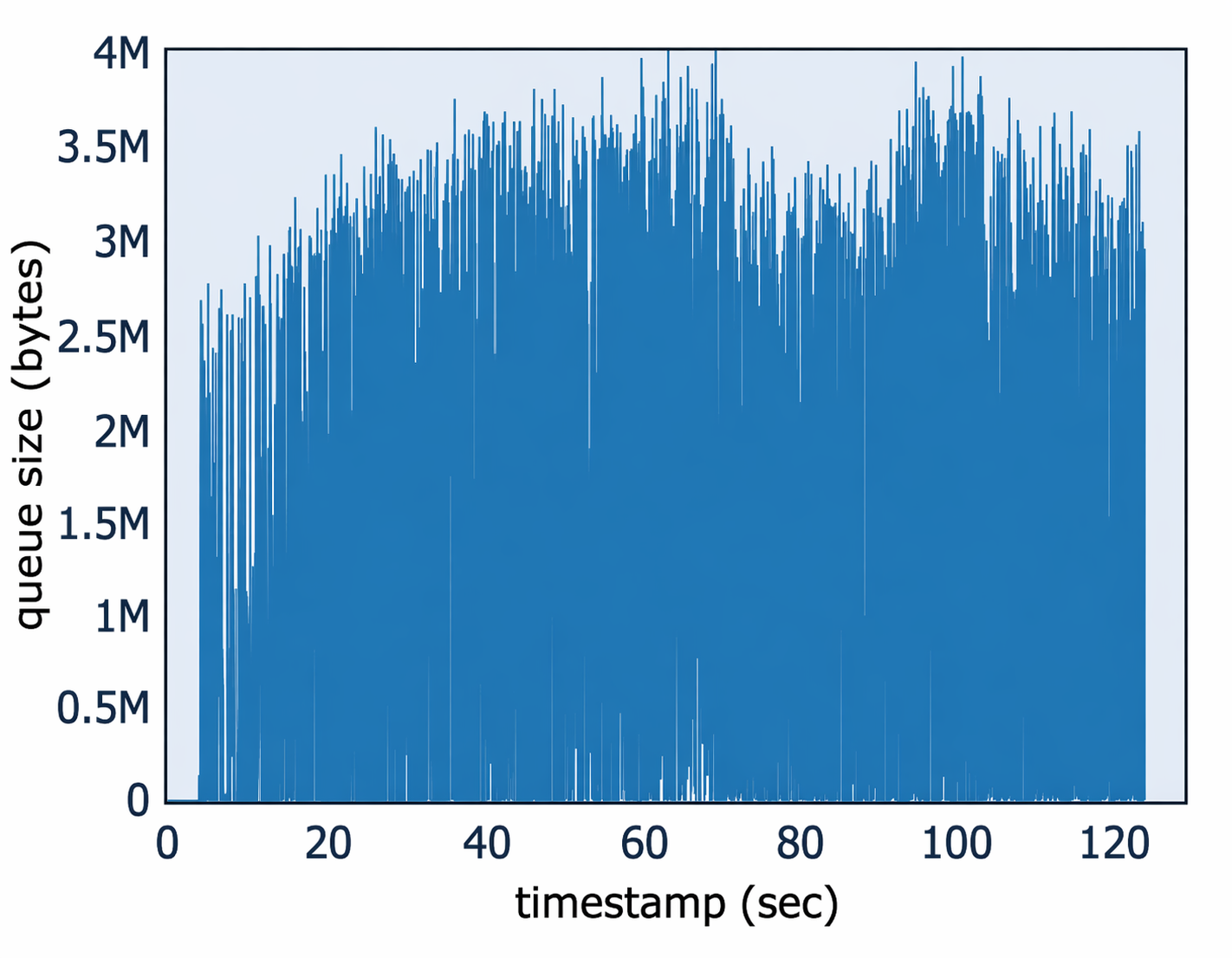}
    \caption{More aggressive CC profile}
    \label{fig:dcqcn-more}
  \end{subfigure}
  \begin{subfigure}{0.33\textwidth}
    \includegraphics[width=\textwidth]{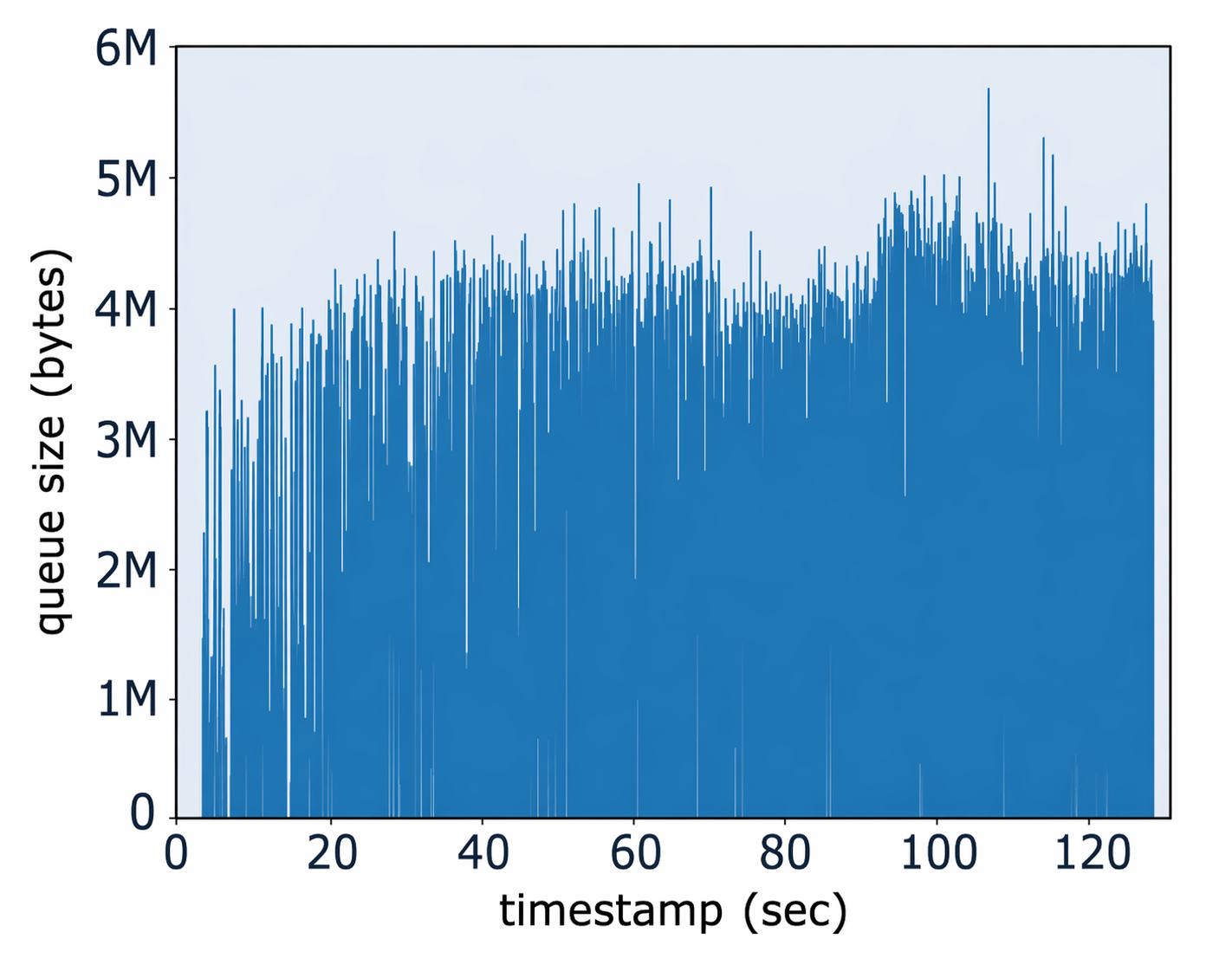}
    \caption{Most aggressive CC profile}
    \label{fig:dcqcn-most}
  \end{subfigure}
  \caption{DCQCN leaf-host queue dynamics in a 15:1 incast where flows arrive 5s apart.}
\end{figure*}

\noindent\textbf{Load balancing micro-benchmarks.}  MRC's spraying and
active load balancing is designed to eliminate congestion caused by
flow collisions.  We present some results using RoCEv2 that illustrate
the nature of this problem.

We have two racks (8 hosts) send data to the other two racks, in a
one-to-one pattern using \texttt{ib\_write\_bw} that fully loads the T0
uplinks. When using a single QP per transfer with RoCEv2, some flows get line rate, but some suffer from flow
collisions and achieve only a half or a third of line
rate; the outcome is the same when using DCQCN \cite{dcqcn} or PFC
alone and the number of collisions varies between runs. See Figure
\ref{fig:permpfc1qp} for a run with PFC but no DCQCN.  In contrast, in
this same scenario, MRC using a single QP is able to achieve 390~Gbps
for all flows (not shown).

When we use 8QPs per transfer, it matters whether we use PFC alone or DCQCN with PFC, as shown in
Figures~\ref{fig:permpfc8qp} and \ref{fig:permdcqcn8qp}. When only PFC is used,
having multiple QPs only marginally helps: the QPs experiencing most congestion
will cause the sending NIC to be throttled by PFC, which will in turn slow down
all other QPs from the same host. DCQCN avoids this phenomenon by reducing the sending
rate based on ECN signals \cite{dcqcn}. Again, for MRC the throughput is line rate
also with 8QPs.

\noindent\textbf{Collateral damage of incast.}
Lossless networks struggle with incast traffic patterns when the
congestion spreads and can affect unrelated ``victim'' traffic as
discussed in Section \ref{sec:collateral}. Here we present some additional
results for the same cross-spine 7 to 1 incast traffic pattern run in
parallel to a ``victim'' connection when using RoCEv2 with PFC only, and
when using MRC with 8QPs. Results for RoCEv2 with DCQCN and MRC with
1QP have been presented in Figures \ref{fig:incastdcqcn1qp},
\ref{fig:incastdcqcn8qp} and \ref{fig:incastmrc1qp}.

We show the results in Figures \ref{fig:incastmrc8qp},
\ref{fig:incastpfc1qp} and \ref{fig:incastpfc8qp}.
With PFC alone the effect on the victim flow is dire, with the victim
flow only achieving 30 to 100Gbps depending on ECMP path choice and
fairness. MRC with 8QPs behaves exactly the same as with 1QP,
perfectly sharing the bottleneck link among the incast flows and has
no impact on the victim flow.

In principle DCQCN parameter tuning should fix this issue. However,
configuring DCQCN properly is very hard because it is traffic pattern
specific.

To show why this is the case, we show the TOR queue size to the
destination during a 15-to-one incast where flows arrive sequentially every 5s.
We tested three different DCQCN
profiles recommended to customers: the default one
(Fig.~\ref{fig:dcqcn-default}), a more aggressive CC profile
(Fig.~\ref{fig:dcqcn-more}) and a most aggressive one
(Fig.~\ref{fig:dcqcn-most}). In the default case, the queue controlled with
a small number of flows (less than 10) and then it goes into PFC mode,
but the bottleneck throughput is line-rate. In the more aggressive
profile the queue is controlled (with some wild dynamic range) but the
queue is sometimes empty and this leads to around 10\% loss in
bottleneck throughput. Finally, with the most aggressive setting, the
queue usage spikes are similar but average utilization is lower, and
the total throughput is 20\% slower than line-rate.

%% file: biblio.bib
@techreport{mrc-spec,
  author       = {Rip Sohan and Eric Spada and Eric Davis and Mark Handley and Idan Burstein and Tony Hurson and Jithin Jose and Vivek Kashyap and Rong Pan and Sayantan Sur},
  title        = {{Multipath Reliable Connection (MRC) Specification}},
  institution  = {{Open Compute Project}},
  year         = {2026},
  type         = {Specification},
  number       = {Version 1.0},
  howpublished = {\url{https://www.opencompute.org/documents/ocp-mrc-1-0-pdf}}
}

@inproceedings {stardust,
author = {Noa Zilberman and Gabi Bracha and Golan Schzukin},
title = {Stardust: Divide and Conquer in the Data Center Network},
booktitle = {16th USENIX Symposium on Networked Systems Design and Implementation (NSDI 19)},
year = {2019},
isbn = {978-1-931971-49-2},
address = {Boston, MA},
pages = {141--160},
url = {https://www.usenix.org/conference/nsdi19/presentation/zilberman},
publisher = {USENIX Association},
month = feb,
}

@inproceedings{mptcp-dc,
 author = {Raiciu, Costin and Barre, Sebastien and Pluntke, Christopher and Greenhalgh, Adam and Wischik, Damon and Handley, Mark},
 title = {{Improving Datacenter Performance and Robustness with Multipath TCP}},
 booktitle = {Special Interest Group on Data Communication (SIGCOMM)},
 year = {2010},
 publisher = {ACM}
}

@inproceedings{hedera,
 author = {Al-Fares, Mohammad and Radhakrishnan, Sivasankar and Raghavan, Barath and Huang, Nelson and Vahdat, Amin},
 title = {{Hedera: Dynamic Flow Scheduling for Data Center Networks}},
 booktitle = {Networked Systems Design and Implementation (NSDI)},
 year = {2010},
 publisher = {USENIX Association}
}

@inproceedings{conga,
 author = {Alizadeh, Mohammad and Edsall, Tom and Dharmapurikar, Sarang and Vaidyanathan, Ramanan and Chu, Kevin and Fingerhut, Andy and Lam, Vinh The and Matus, Francis and Pan, Rong and Yadav, Navindra and Varghese, George},
 title = {{CONGA: Distributed Congestion-aware Load Balancing for Datacenters}},
 booktitle = {Special Interest Group on Data Communication (SIGCOMM)},
 year = {2014},
 publisher = {ACM}
}

@inproceedings{flowbender,
 author = {Kabbani, Abdul and Vamanan, Balajee and Hasan, Jahangir and Duchene, Fabien},
 title = {{FlowBender: Flow-level Adaptive Routing for Improved Latency and Throughput in Datacenter Networks}},
 booktitle = {Conference on Emerging Networking Experiments and Technologies (CoNEXT)},
 year = {2014},
 publisher = {ACM}
}

@inproceedings{spray,
 author = {Dixit, Advait and Prokash, Pawan and Hu, Charlie Y. and Kompella, Ramona R}, 
 booktitle = {International Conference on Computer Communications (INFOCOM)}, 
 title = {{On the Impact of Packet Spraying in Data Center Networks}}, 
 year = {2013}, 
 publisher = {IEEE}
}

@inproceedings{homa,
 author = {Montazeri, Behnam and Li, Yilong and Alizadeh, Mohammad and Ousterhout, John},
 title = {{Homa: A Receiver-driven Low-latency Transport Protocol Using Network Priorities}},
 booktitle = {Special Interest Group on Data Communication (SIGCOMM)},
 year = {2018},
 publisher = {ACM}
}

@inproceedings{ndp,
 author = {Handley, Mark and Raiciu, Costin and Agache, Alexandru and Voinescu, Andrei and Moore, Andrew W. and Antichi, Gianni and W\'{o}jcik, Marcin},
 title = {{Re-architecting Datacenter Networks and Stacks for Low Latency and High Performance}},
 booktitle = {Special Interest Group on Data Communication (SIGCOMM)},
 year = {2017},
 publisher = {ACM}
}

@inproceedings{IRN,
author = {Mittal, Radhika and Shpiner, Alexander and Panda, Aurojit and Zahavi, Eitan and Krishnamurthy, Arvind and Ratnasamy, Sylvia and Shenker, Scott},
title = {Revisiting Network Support for {RDMA}},
year = {2018},
isbn = {9781450355674},
publisher = {Association for Computing Machinery},
address = {New York, NY, USA},
url = {https://doi.org/10.1145/3230543.3230557},
doi = {10.1145/3230543.3230557},
booktitle = {Proceedings of the 2018 Conference of the ACM Special Interest Group on Data Communication},
pages = {313–326},
numpages = {14},
location = {Budapest, Hungary},
series = {SIGCOMM '18}
}

@inproceedings {mp-rdma,
author = {Yuanwei Lu and Guo Chen and Bojie Li and Kun Tan and Yongqiang Xiong and Peng Cheng and Jiansong Zhang and Enhong Chen and Thomas Moscibroda},
title = {Multi-Path Transport for {RDMA} in Datacenters},
booktitle = {15th {USENIX} Symposium on Networked Systems Design and Implementation ({NSDI} 18)},
year = {2018},
isbn = {978-1-939133-01-4},
address = {Renton, WA},
pages = {357--371},
url = {https://www.usenix.org/conference/nsdi18/presentation/lu},
publisher = {{USENIX} Association},
month = apr
}

@inproceedings{dcqcn,
author = {Zhu, Yibo and Eran, Haggai and Firestone, Daniel and Guo, Chuanxiong and Lipshteyn, Marina and Liron, Yehonatan and Padhye, Jitendra and Raindel, Shachar and Yahia, Mohamad Haj and Zhang, Ming},
title = {Congestion Control for Large-Scale {RDMA} Deployments},
year = {2015},
isbn = {9781450335423},
publisher = {Association for Computing Machinery},
address = {New York, NY, USA},
url = {https://doi.org/10.1145/2785956.2787484},
doi = {10.1145/2785956.2787484},
booktitle = {Proceedings of the 2015 ACM Conference on Special Interest Group on Data Communication},
pages = {523–536},
numpages = {14},
location = {London, United Kingdom},
series = {SIGCOMM '15}
}

@Misc{IBTA:roce,
  author = 	 {{InfiniBand Trade Association (IBTA)}},
  title = 	 {The {RoCE} Initiative},
  note = 	 {(Accessed: May 2021)},
  url = {https://www.infinibandta.org/roce-initiative/}
}

@misc{dlb,
      title={{ECMP Dynamic Load Balancing}},
      author={Broadcom},
      year={2019},
      howpublished = {\url{https://docs.broadcom.com/doc/56980-DS}}
}

@inproceedings{falcon,
author = {Singhvi, Arjun and Dukkipati, Nandita and Chandra, Prashant and Wassel, Hassan M. G. and Sharma, Naveen Kr. and Rebello, Anthony and Schuh, Henry and Kumar, Praveen and Montazeri, Behnam and Bansod, Neelesh and Thomas, Sarin and Cho, Inho and Seibert, Hyojeong Lee and Wu, Baijun and Yang, Rui and Li, Yuliang and Huang, Kai and Yin, Qianwen and Agarwal, Abhishek and Vaduvatha, Srinivas and Wang, Weihuang and Moshref, Masoud and Ji, Tao and Wetherall, David and Vahdat, Amin},
title = {Falcon: A Reliable, Low Latency Hardware Transport},
year = {2025},
isbn = {9798400715242},
publisher = {Association for Computing Machinery},
address = {New York, NY, USA},
url = {https://doi.org/10.1145/3718958.3754353},
doi = {10.1145/3718958.3754353},
booktitle = {Proceedings of the ACM SIGCOMM 2025 Conference},
pages = {248–263},
numpages = {16},
location = {S\~{a}o Francisco Convent, Coimbra, Portugal},
series = {SIGCOMM '25}
}

@misc{llama3,
      title={The {LLaMa} 3 Herd of Models}, 
      author={Aaron Grattafiori et al},
      year={2024},
      eprint={2407.21783},
      archivePrefix={arXiv},
      primaryClass={cs.AI},
      url={https://arxiv.org/abs/2407.21783}, 
}

@inproceedings{meta-roce,
author = {Gangidi, Adithya and Miao, Rui and Zheng, Shengbao and Bondu, Sai Jayesh and Goes, Guilherme and Morsy, Hany and Puri, Rohit and Riftadi, Mohammad and Shetty, Ashmitha Jeevaraj and Yang, Jingyi and Zhang, Shuqiang and Fernandez, Mikel Jimenez and Gandham, Shashidhar and Zeng, Hongyi},
title = {RDMA over Ethernet for Distributed Training at Meta Scale},
year = {2024},
isbn = {9798400706141},
publisher = {Association for Computing Machinery},
address = {New York, NY, USA},
url = {https://doi.org/10.1145/3651890.3672233},
doi = {10.1145/3651890.3672233},
booktitle = {Proceedings of the ACM SIGCOMM 2024 Conference},
pages = {57–70},
numpages = {14},
location = {Sydney, NSW, Australia},
series = {ACM SIGCOMM '24}
}

@misc{uet,
      title={{Ultra Ethernet} Specification v1.0.1},
      author={Ultra Ethernet Consortium},
      year={2025},
      url={https://ultraethernet.org/wp-content/uploads/sites/20/2025/10/UE-Specification-1.0.1.pdf}
}

@misc{p4,
      title={{P4 Open Source Programming Language}},
      howpublished = {\url{https://p4.org/}}
}

@misc{rfc8986,
  series = {Request for Comments},
  number = {8986},
  howpublished = {RFC 8986},
  publisher = {RFC Editor},
  doi = {10.17487/RFC8986},
  url = {https://www.rfc-editor.org/info/rfc8986},
  author = {Clarence Filsfils and Darren Dukes and Stefano Previdi and John Leddy and Satoru Matsushima and Daniel Voyer},
  title = {{Segment Routing over IPv6 (SRv6) Network Programming}},
  year = {2021},
  month = feb,
}

@misc{rfc9800,
  series = {Request for Comments},
  number = {9800},
  howpublished = {RFC 9800},
  publisher = {RFC Editor},
  doi = {10.17487/RFC9800},
  url = {https://www.rfc-editor.org/info/rfc9800},
  author = {Weiqiang Cheng and Clarence Filsfils and Zhenbin Li and Bruno Decraene and Dezhong Cai and Daniel Voyer and Francois Clad and Shay Zadok and Jim Guichard and Aihua Liu and Robert Raszuk and Cheng Li},
  title = {{Compressed SRv6 Segment List Encoding}},
  year = {2025},
  month = jun,
}

@article{switchfailures,
author = {Singh, Rachee and Mukhtar, Muqeet and Krishna, Ashay and Parkhi, Aniruddha and Padhye, Jitendra and Maltz, David},
title = {Surviving switch failures in cloud datacenters},
year = {2021},
issue_date = {April 2021},
publisher = {Association for Computing Machinery},
address = {New York, NY, USA},
volume = {51},
number = {2},
issn = {0146-4833},
url = {https://doi.org/10.1145/3464994.3464996},
doi = {10.1145/3464994.3464996},
journal = {SIGCOMM Comput. Commun. Rev.},
month = may,
pages = {2–9},
numpages = {8}
}

@article{bugs-os-nos,
author = {Yin, Zuoning and Caesar, Matthew and Zhou, Yuanyuan},
title = {Towards understanding bugs in open source router software},
year = {2010},
issue_date = {July 2010},
publisher = {Association for Computing Machinery},
address = {New York, NY, USA},
volume = {40},
number = {3},
issn = {0146-4833},
url = {https://doi.org/10.1145/1823844.1823849},
doi = {10.1145/1823844.1823849},
journal = {SIGCOMM Comput. Commun. Rev.},
month = jun,
pages = {34–40},
numpages = {7}
}

@inproceedings{crystalnet,
author = {Liu, Hongqiang and Zhu, Yibo and Padhye, Jitu and Cao, Jiaxin and Tallapragada,  Sri and Lopes, Nuno and Rybalchenko, Andrey and Lu, Guohan and Yuan, Lihua},
title = {{CrystalNet}: Faithfully Emulating Large Production Networks},
booktitle = {SOSP '17 Proceedings of the 26th Symposium on Operating Systems Principles},
year = {2017},
month = {October},
publisher = {ACM},
url = {https://www.microsoft.com/en-us/research/publication/crystalnet-faithfully-emulating-large-production-networks/},
pages = {599-613},
isbn = {978-1-4503-5085-3},
edition = {SOSP '17 Proceedings of the 26th Symposium on Operating Systems Principles},
}

@article{pingmesh,
author = {Guo, Chuanxiong and Yuan, Lihua and Xiang, Dong and Dang, Yingnong and Huang, Ray and Maltz, Dave and Liu, Zhaoyi and Wang, Vin and Pang, Bin and Chen, Hua and Lin, Zhi-Wei and Kurien, Varugis},
title = {Pingmesh: A Large-Scale System for Data Center Network Latency Measurement and Analysis},
year = {2015},
issue_date = {October 2015},
publisher = {Association for Computing Machinery},
address = {New York, NY, USA},
volume = {45},
number = {4},
issn = {0146-4833},
url = {https://doi.org/10.1145/2829988.2787496},
doi = {10.1145/2829988.2787496},
journal = {SIGCOMM Comput. Commun. Rev.},
month = aug,
pages = {139–152},
numpages = {14}
}

@inproceedings{hpn,
author = {Qian, Kun and Xi, Yongqing and Cao, Jiamin and Gao, Jiaqi and Xu, Yichi and Guan, Yu and Fu, Binzhang and Shi, Xuemei and Zhu, Fangbo and Miao, Rui and Wang, Chao and Wang, Peng and Zhang, Pengcheng and Zeng, Xianlong and Ruan, Eddie and Yao, Zhiping and Zhai, Ennan and Cai, Dennis},
title = {Alibaba {HPN}: A Data Center Network for Large Language Model Training},
year = {2024},
isbn = {9798400706141},
publisher = {Association for Computing Machinery},
address = {New York, NY, USA},
url = {https://doi.org/10.1145/3651890.3672265},
doi = {10.1145/3651890.3672265},
booktitle = {Proceedings of the ACM SIGCOMM 2024 Conference},
pages = {691–706},
numpages = {16},
location = {Sydney, NSW, Australia},
series = {ACM SIGCOMM '24}
}

@INPROCEEDINGS{railonly,
  author={Wang, Weiyang and Ghobadi, Manya and Shakeri, Kayvon and Zhang, Ying and Hasani, Naader},
  booktitle={2024 IEEE Symposium on High-Performance Interconnects (HOTI)}, 
  title={Rail-only: A Low-Cost High-Performance Network for Training {LLMs} with Trillion Parameters}, 
  year={2024},
  volume={},
  number={},
  pages={1-10},
  doi={10.1109/HOTI63208.2024.00013}}

@inproceedings{99probs,
author = {Gherghescu, Alexandru M. and B\u{a}doiu, Vlad-Andrei and Agache, Alexandru and Dumitru, Mihai-Valentin and Vasilescu, Iuliu and Mantu, Radu and Raiciu, Costin},
title = {I've Got 99 Problems But {FLOPS} Ain't One},
year = {2024},
isbn = {9798400712722},
publisher = {Association for Computing Machinery},
address = {New York, NY, USA},
url = {https://doi.org/10.1145/3696348.3696893},
doi = {10.1145/3696348.3696893},
booktitle = {Proceedings of the 23rd ACM Workshop on Hot Topics in Networks},
pages = {195–204},
numpages = {10},
location = {Irvine, CA, USA},
series = {HotNets '24}
}

@inproceedings{filsfils_srv6_ai,
  author = {Clarence Filsfils and Pablo Camarillo and Ahmed Abdelsalam and Arianna Quinci and Angelo Tulumello and Andrea Mayer and Pierpaolo Loreti and Lorenzo Bracciale and Stefano Salsano},
  title = {Toward Deterministic Path Placement in {AI} Backends: A Practical {SRv6}-Based Architecture},
  booktitle = {21st International Conference on Network and Service Management (CNSM)},
  year = {2025},
  address = {Bologna, Italy},
  month = oct,
  isbn = {978-3-903176-75-1},
  publisher = {IFIP},
  url = {https://dl.ifip.org/db/conf/cnsm/cnsm2025/1571173126.pdf}
}

@article{10.1145/3320060,
author = {Ben-Nun, Tal and Hoefler, Torsten},
title = {Demystifying Parallel and Distributed Deep Learning: An In-depth Concurrency Analysis},
year = {2019},
issue_date = {July 2020},
publisher = {Association for Computing Machinery},
address = {New York, NY, USA},
volume = {52},
number = {4},
issn = {0360-0300},
url = {https://doi.org/10.1145/3320060},
doi = {10.1145/3320060},
journal = {ACM Comput. Surv.},
month = aug,
articleno = {65},
numpages = {43}
}

@misc{rajbhandari2022,
      title={{DeepSpeed-MoE}: Advancing Mixture-of-Experts Inference and Training to Power Next-Generation AI Scale}, 
      author={Samyam Rajbhandari and Conglong Li and Zhewei Yao and Minjia Zhang and Reza Yazdani Aminabadi and Ammar Ahmad Awan and Jeff Rasley and Yuxiong He},
      year={2022},
      eprint={2201.05596},
      archivePrefix={arXiv},
      primaryClass={cs.LG},
      url={https://arxiv.org/abs/2201.05596}, 
}

@article{yan2026scalable,
  title={Scalable Training of Mixture-of-Experts Models with {Megatron Core}},
  author={Yan, Zijie and Bai, Hongxiao and Yao, Xin and Liu, Dennis and Liu, Tong and Liu, Hongbin and Li, Pingtian and Wu, Evan and Fan, Shiqing and Tao, Li and others},
  journal={arXiv preprint arXiv:2603.07685},
  year={2026}
}

@article{10.1145/2408776.2408794,
author = {Dean, Jeffrey and Barroso, Luiz Andr\'{e}},
title = {The tail at scale},
year = {2013},
issue_date = {February 2013},
publisher = {Association for Computing Machinery},
address = {New York, NY, USA},
volume = {56},
number = {2},
issn = {0001-0782},
url = {https://doi.org/10.1145/2408776.2408794},
doi = {10.1145/2408776.2408794},
journal = {Commun. ACM},
month = feb,
pages = {74–80},
numpages = {7}
}

@inproceedings{10.1109/SC.2010.12,
author = {Hoefler, Torsten and Schneider, Timo and Lumsdaine, Andrew},
title = {Characterizing the Influence of System Noise on Large-Scale Applications by Simulation},
year = {2010},
isbn = {9781424475599},
publisher = {IEEE Computer Society},
address = {USA},
url = {https://doi.org/10.1109/SC.2010.12},
doi = {10.1109/SC.2010.12},
booktitle = {Proceedings of the 2010 ACM/IEEE International Conference for High Performance Computing, Networking, Storage and Analysis},
pages = {1–11},
numpages = {11},
series = {SC '10}
}

@inproceedings{10.1145/1048935.1050204,
author = {Petrini, Fabrizio and Kerbyson, Darren J. and Pakin, Scott},
title = {The Case of the Missing Supercomputer Performance: Achieving Optimal Performance on the 8,192 Processors of {ASCI Q}},
year = {2003},
isbn = {1581136951},
publisher = {Association for Computing Machinery},
address = {New York, NY, USA},
url = {https://doi.org/10.1145/1048935.1050204},
doi = {10.1145/1048935.1050204},
booktitle = {Proceedings of the 2003 ACM/IEEE Conference on Supercomputing},
pages = {55},
location = {Phoenix, AZ, USA},
series = {SC '03}
}

@misc{ue-overview,
      title={Ultra Ethernet's Design Principles and Architectural Innovations}, 
      author={Torsten Hoefler and Karen Schramm and Eric Spada and Keith Underwood and Cedell Alexander and Bob Alverson and Paul Bottorff and Adrian Caulfield and Mark Handley and Cathy Huang and Costin Raiciu and Abdul Kabbani and Eugene Opsasnick and Rong Pan and Adee Ran and Rip Sohan},
      year={2025},
      eprint={2508.08906},
      archivePrefix={arXiv},
      primaryClass={cs.NI},
      url={https://arxiv.org/abs/2508.08906}, 
}

@inproceedings{microTE,
author = {Benson, Theophilus and Anand, Ashok and Akella, Aditya and Zhang, Ming},
title = {MicroTE: fine grained traffic engineering for data centers},
year = {2011},
isbn = {9781450310413},
publisher = {Association for Computing Machinery},
address = {New York, NY, USA},
url = {https://doi.org/10.1145/2079296.2079304},
doi = {10.1145/2079296.2079304},
booktitle = {Proceedings of the Seventh COnference on Emerging Networking EXperiments and Technologies},
articleno = {8},
numpages = {12},
location = {Tokyo, Japan},
series = {CoNEXT '11}
}

@inproceedings {201562,
author = {Erico Vanini and Rong Pan and Mohammad Alizadeh and Parvin Taheri and Tom Edsall},
title = {Let It Flow: Resilient Asymmetric Load Balancing with Flowlet Switching},
booktitle = {14th USENIX Symposium on Networked Systems Design and Implementation (NSDI 17)},
year = {2017},
isbn = {978-1-931971-37-9},
address = {Boston, MA},
pages = {407--420},
url = {https://www.usenix.org/conference/nsdi17/technical-sessions/presentation/vanini},
publisher = {USENIX Association},
month = mar
}

@ARTICLE{11299491,
  author={Bonato, Tommaso and De Sensi, Daniele and Di Girolamo, Salvatore and Bataineh, Abdulla and Hewson, David and Roweth, Duncan and Hoefler, Torsten},
  journal={IEEE Transactions on Networking}, 
  title={Flowcut Switching: High-Performance Adaptive Routing With In-Order Delivery Guarantees}, 
  year={2026},
  volume={34},
  number={},
  pages={1974-1987},
  doi={10.1109/TON.2025.3636209}}

@inproceedings{flowcell,
author = {He, Keqiang and Rozner, Eric and Agarwal, Kanak and Felter, Wes and Carter, John and Akella, Aditya},
title = {Presto: Edge-Based Load Balancing for Fast Datacenter Networks},
year = {2015},
isbn = {9781450335423},
publisher = {Association for Computing Machinery},
address = {New York, NY, USA},
url = {https://doi.org/10.1145/2785956.2787507},
doi = {10.1145/2785956.2787507},
booktitle = {Proceedings of the 2015 ACM Conference on Special Interest Group on Data Communication},
pages = {465–478},
numpages = {14},
location = {London, United Kingdom},
series = {SIGCOMM '15}
}

@inproceedings{10.1145/3544216.3544226,
author = {Qureshi, Mubashir Adnan and Cheng, Yuchung and Yin, Qianwen and Fu, Qiaobin and Kumar, Gautam and Moshref, Masoud and Yan, Junhua and Jacobson, Van and Wetherall, David and Kabbani, Abdul},
title = {{PLB}: congestion signals are simple and effective for network load balancing},
year = {2022},
isbn = {9781450394208},
publisher = {Association for Computing Machinery},
address = {New York, NY, USA},
url = {https://doi.org/10.1145/3544216.3544226},
doi = {10.1145/3544216.3544226},
booktitle = {Proceedings of the ACM SIGCOMM 2022 Conference},
pages = {207–218},
numpages = {12},
location = {Amsterdam, Netherlands},
series = {SIGCOMM '22}
}

@inproceedings{drill,
author = {Ghorbani, Soudeh and Yang, Zibin and Godfrey, P. Brighten and Ganjali, Yashar and Firoozshahian, Amin},
title = {{DRILL}: Micro Load Balancing for Low-Latency Data Center Networks},
year = {2017},
isbn = {9781450346535},
publisher = {Association for Computing Machinery},
address = {New York, NY, USA},
url = {https://doi.org/10.1145/3098822.3098839},
doi = {10.1145/3098822.3098839},
booktitle = {Proceedings of the Conference of the ACM Special Interest Group on Data Communication},
pages = {225–238},
numpages = {14},
location = {Los Angeles, CA, USA},
series = {SIGCOMM '17}
}

@misc{nccl,
      title={Demystifying {NCCL}: An In-depth Analysis of {GPU} Communication Protocols and Algorithms}, 
      author={Zhiyi Hu and Siyuan Shen and Tommaso Bonato and Sylvain Jeaugey and Cedell Alexander and Eric Spada and James Dinan and Jeff Hammond and Torsten Hoefler},
      year={2026},
      eprint={2507.04786},
      archivePrefix={arXiv},
      primaryClass={cs.DC},
      url={https://arxiv.org/abs/2507.04786}, 
}

@inproceedings{hermes,
author = {Zhang, Hong and Zhang, Junxue and Bai, Wei and Chen, Kai and Chowdhury, Mosharaf},
title = {Resilient Datacenter Load Balancing in the Wild},
year = {2017},
isbn = {9781450346535},
publisher = {Association for Computing Machinery},
address = {New York, NY, USA},
url = {https://doi.org/10.1145/3098822.3098841},
doi = {10.1145/3098822.3098841},
booktitle = {Proceedings of the Conference of the ACM Special Interest Group on Data Communication},
pages = {253–266},
numpages = {14},
location = {Los Angeles, CA, USA},
series = {SIGCOMM '17}
}

@misc{ecmp,
        series =        {Request for Comments},
        number =        2992,
        howpublished =  {RFC 2992},
        publisher =     {RFC Editor},
        url =           {https://www.ietf.org/rfc/rfc2992.txt},
        author =        {C. Hopps},
        title =         {{Analysis of an Equal-Cost Multi-Path Algorithm}},
        year =          2009,
        month =         nov,
}

@misc{reps,
      title={{REPS}: Recycled Entropy Packet Spraying for Adaptive Load Balancing and Failure Mitigation}, 
      author={Tommaso Bonato and Abdul Kabbani and Ahmad Ghalayini and Michael Papamichael and Mohammad Dohadwala and Lukas Gianinazzi and Mikhail Khalilov and Elias Achermann and Daniele De Sensi and Torsten Hoefler},
      year={2026},
      eprint={2407.21625},
      archivePrefix={arXiv},
      primaryClass={cs.NI},
      doi={https://doi.org/10.1145/3767295.3769320},
      url={https://arxiv.org/abs/2407.21625}, 
}

@misc{ncclx,
      title={Collective Communication for 100k+ {GPUs}}, 
      author={Min Si and Pavan Balaji and Yongzhou Chen and Ching-Hsiang Chu and Adi Gangidi and Saif Hasan and Subodh Iyengar and Dan Johnson and Bingzhe Liu and Regina Ren and Deep Shah and Ashmitha Jeevaraj Shetty and Greg Steinbrecher and Yulun Wang and Bruce Wu and Xinfeng Xie and Jingyi Yang and Mingran Yang and Kenny Yu and Minlan Yu and Cen Zhao and Wes Bland and Denis Boyda and Suman Gumudavelli and Prashanth Kannan and Cristian Lumezanu and Rui Miao and Zhe Qu and Venkat Ramesh and Maxim Samoylov and Jan Seidel and Srikanth Sundaresan and Feng Tian and Qiye Tan and Shuqiang Zhang and Yimeng Zhao and Shengbao Zheng and Art Zhu and Hongyi Zeng},
      year={2026},
      eprint={2510.20171},
      archivePrefix={arXiv},
      primaryClass={cs.DC},
      url={https://arxiv.org/abs/2510.20171}, 
}

@misc{msccl,
      title={{TACCL}: Guiding Collective Algorithm Synthesis using Communication Sketches}, 
      author={Aashaka Shah and Vijay Chidambaram and Meghan Cowan and Saeed Maleki and Madan Musuvathi and Todd Mytkowicz and Jacob Nelson and Olli Saarikivi and Rachee Singh},
      year={2022},
      eprint={2111.04867},
      archivePrefix={arXiv},
      primaryClass={cs.DC},
      url={https://arxiv.org/abs/2111.04867}, 
}

@article{uccl2,
  title={{UCCL-Tran}: An Extensible Software Transport Layer for Machine Learning Workloads},
  author={Zhou, Yang and Chen, Zhongjie and Mao, Ziming and Lao, ChonLam and Yang, Shuo and Kannan, Pravein Govindan and Gao, Jiaqi and Zhao, Yilong and Wu, Yongji and You, Kaichao and Ren, Fengyuan and Xu, Zhiying and Raiciu, Costin and Stoica, Ion},
  journal={USENIX OSDI},
  year={2026}
}
